\RequirePackage{fix-cm}
\documentclass[smallcondensed]{svjour3}     
\smartqed  
\usepackage{graphicx}
\usepackage[numbers,square,sort]{natbib}
\usepackage{mathptmx}      
\journalname{Journal of Scientific Computing}
\begin{document}

\title{Numerical treatment of interfaces for second-order wave equations
}


\author{Florencia Parisi         \and
        Mariana C\'ecere         \and
        Mirta Iriondo            \and
        Oscar Reula
}


\institute{F. Parisi \and M. C\'ecere \and M. Iriondo \and O. Reula \at
              Instituto de F\'{\i}sica Enrique Gaviola, CONICET, FaMAF - Universidad Nacional de C\'ordoba, Ciudad Universitaria, 5000 C\'ordoba, Argentina \\
              \email{fparisi@famaf.unc.edu.ar}           
           \and
           M. C\'ecere \at
              \email{cecere@famaf.unc.edu.ar}
                         \and
           M. Iriondo \at
              \email{mirta@famaf.unc.edu.ar}
              \and
           O. Reula \at
              \email{reula@famaf.unc.edu.ar}
    \and
           M. C\'ecere \at
            Instituto de Astronom\'{\i}a Te\'orica y Experimental, CONICET-UNC, Observatorio Astron\'omico de C\'ordoba, X5000BGR,  C\'ordoba, Argentina \\
              \email{cecere@famaf.unc.edu.ar}
                         \and
}

\date{Received: date / Accepted: June 11, 2014}

\maketitle

\begin{abstract}

In this article we develop a numerical scheme to deal with interfaces between touching numerical grids when solving the
second-order wave equation.
We show that it is possible to implement an interface scheme of  ``penalty'' type for the second-order wave equation,
similar to the ones
used for first-order hyperbolic and parabolic equations, and the second-order scheme used by Mattsson et al. \cite{Mattsson-Ham}.
These schemes, known as SAT schemes for finite difference approximations and penalties for spectral ones, and ours share similar properties but in our case  one needs to pass at the interface a smaller
 amount of data than   previously known schemes.
This is important for multi-block parallelizations in several dimensions, for it implies that one obtains the same solution
quality while sharing among different computational grids only a  fraction of the data one would need for a comparable
(in accuracy) SAT or Mattsson et al.'s scheme.



The semi-discrete approximation used here preserves the norm and uses standard finite-difference operators satisfying summation
by parts.
For the time integrator we use a semi-implicit IMEX Runge-Kutta method. This is crucial, since the explicit Runge-Kutta method
would be impractical given the severe restrictions
that arise from the stiff parts of the equations.

\end{abstract}

\section{Introduction}
\label{intro}

In many situations modern simulations of physical models based on systems of partial differential equations require the use of several grid patches.
This could be because the topology of the underlying space is not trivial and so cannot be described by a unique chart (as in the case of codes that evolve fields on
a sphere) or it could be because the problem is too large for a single computer and so it has to be partitioned  to be solved parallely in clusters of computers, or because some areas of the integration domain need more resolution than others.
Information needs to be passed among these grids in a dynamical and synchronized fashion. Therefore it is important to devise methods that guarantee the stability
of the global solution and that require the minimal amount of information to be transferred at grid interfaces, preserving this way a given accuracy.

For some years now, several numerical techniques have been available to deal with these interface problems when solving first-order hyperbolic or parabolic equations.
Some of them use interpolation between overlapping regions, while others use penalties which modify the system at boundary grid points by including information
from the same space points at neighboring grids \cite{Carpenter1999341}.
This last method is preferable in many situations, for it has very nice properties. The most interesting one is the fact that it is constructed so that the resulting semi-discrete system preserves the corresponding continuum energy estimate of its constant-coefficient linear system. Thus we can ensure that, at least for linear constant-coefficient systems, the scheme is stable.

Another property that makes these schemes attractive is that the amount of information one has to pass from one grid to the next is minimal.
Thus these schemes are optimal for massive parallel computations, where calculations must be split among several CPU/GPU's, and communication
among them usually would add a non-trivial overhead to the computation of the solution.

Furthermore,  the accuracy of the method is only determined by the precision order of the time and space operators. Hence the amount of data transferred is constant. This is not the case though when interpolation is used since in order to increase the accuracy of that scheme
one has to take more points on each of the grids and use higher-order interpolants, leading to a larger amount of data that needs to be transmitted.

For a second-order hyperbolic equation these methods can be used, but with the drawback that one must first rewrite it as a
first-order system to be able to apply the approach, by this
creating many more variables and constraints, all of which have to be monitored during the evolution.
This not only increases the required machine memory, but also increases the amount of traffic among processors, due to the
corresponding increase of information that needs to be passed along the boundaries.

In 2008 Mattsson et al. \cite{Mattsson-Ham} developed a method to evolve second-order hyperbolic equations in the same spirit of the standard SAT scheme without the need to reduce it to first-order.
In addition, one of us \cite{reula2011} recently found a way to implement similar techniques for Schr\"odinger's equation. We show in this paper that the same underlying idea in \cite{reula2011} can be extended to also deal
with the second-order wave equation. In this scheme only one field needs to be passed across the boundaries, namely the time derivative of the field, making this a very efficient and simple algorithm, while in the one developed by Mattsson et al., penalties are imposed on the field and its spatial and time derivatives and are present in the evolution of the internal grid points.

The idea of  our scheme is to consider  the boundary
as a repeater which absorbs the incoming wave on one side of the interface and creates an outgoing wave on the other, with a  non-increasing total energy.

This paper is organized as follows: In section 2 we describe the new numerical scheme, and derive the interface terms that need to be added to the equations for this method to work.
There are two different types of  interface terms: one, including only the values of the fields at the same side of the interface (necessary for
cancelling the usual boundary term from the elliptic part of the operator in the energy estimate) and another  one, that can be
regarded as an interaction between the fields at both sides of the interface, and can be considered a penalty, in the sense that it depends on the difference of
fields on both sides of the grids and that it drives  the difference exponentially to zero. The latter is highly stiff, so a semi-implicit
method is needed in order to avoid paying a huge price in the time step. For that purpose, we use IMEX Runge-Kutta methods. Furthermore we present, adapted to our case, the method developed by Mattsson et al.\cite{Mattsson-Ham}.

In section 3 we present some numerical results using the new scheme herein presented.
We compare the results of evolving a one-dimensional system on a circle, first with periodic boundary conditions in a single
grid and then using the
interface scheme between the first and final point of the grid.
We also compare our scheme with the usual SAT approach, in which the system is treated in its first-order form and with the  second order method developed by Mattsson et al.
In addition, we study an implementation of our scheme on a two-dimensional torus.
We analyze convergence, accuracy and stability of the proposed method for different scenarios.

In section 4 we present two applications of our method in much more demanding situations, namely, an equation with
variable coefficients both in space and time which is often used in general relativity as a standard test, and the propagation of a wave on the surface of 2-sphere which has been partitioned into six square grids.

Finally, in section 5 we present our conclusions.

\section{Numerical scheme}
\label{numericalscheme}

Let us consider, for simplicity, a one dimensional problem; the generalization to more dimensions being trivial.
Consider a field $\Phi(x,t): S^{1} \times \Re \to \Re$ satisfying the wave equation:
\begin{equation}
\label{wave}\partial^2_t\Phi=\partial^2_x\Phi,
\end{equation}
and assume sufficiently smooth initial data is given at $t=0$: $\Phi(x,0) = \Phi_{0}(x)$, $\partial_t\Phi(x,0)= \Pi_{0}(x)$.

As mentioned, the traditional way to solve this equation when interfaces are present is by reducing it to a first-order form (hereafter $FO$-{\sl scheme}) by introducing the variables $\Pi:=\partial_t\Phi$ and $\Psi:=\partial_x\Phi$. Then equation (\ref{wave}) is equivalent to the system
\begin{eqnarray}
\partial_{t} \Phi &=& \Pi, \nonumber \\
\partial_{t} \Psi   &=& \partial_x\Pi,\nonumber\\
\label{FO_system}\partial_{t} \Pi   &=& \partial_x\Psi.
\end{eqnarray}
This way of solving the equation has the previously mentioned disadvantage of introducing auxiliary variables, something that can be very expensive in terms of memory, especially when considering systems of wave equations in many dimensions, as is often the case, for instance, in General Relativity.

Furthermore, the use of first-order systems results in less accurate numerical approximations, since the phase error is larger than when using schemes based on the second-order version of the systems as per \cite{kreiss-ortiz}.

We are interested here in solving equation (\ref{wave}) in second-order form for the spatial operators, although we shall keep the first-order form for the time integration, since we will use either a Runge-Kutta or IMEX scheme to advance the fields in time. We therefore consider the system
\begin{eqnarray}
\partial_{t} \Phi &=& \Pi, \nonumber\\
\partial_{t} \Pi   &=& \partial^2_x\Phi,
\label{secondorder}
\end{eqnarray}
and develop a numerical method for solving it when interfaces are present. From here on we deal exclusively with interfaces of the type called {\sl conforming  grids}, in which for each interface point of a grid there corresponds another point from a neighboring grid representing the same spatial point.

Standard theorems guarantee the existence of a solution to equation (\ref{secondorder}) satisfying  the energy norm
\begin{equation}
\label{Energy}\mathcal{E}:= \int \{\Pi^{2} + \nabla \Phi \cdot \nabla \Phi \} dV.
\end{equation}
What we want is to develop a scheme that will preserve the analogous discrete-energy norm, thus guaranteeing stability.

In order to solve this system  we consider the domain that consists of the interval $[0,2]$, where the first and last points are identified, resulting in a circle of length two, with the interface defined in the touching extreme points. For the numerical solution we take a uniformly spaced grid and  we write the discrete solution as a vector
$\{\Phi_{j}\}$, $j=0\dots N$ corresponding to points $x_{j} = dx*j$ with the interspace
between neighboring grid points $dx := \frac{2}{N}$, so that the last point $x_N$ coincides with the
first one $x_0$ at the interface.  With this simplification we focus on the interface treatment avoiding the treatment of boundary conditions.

We introduce the discrete $l^{2}$-norm in the usual fashion,
\begin{equation}
\label{l2_norm}
<\Psi,\Phi> := dx \sum_{j=0}^{N} \sigma_{j}{\Psi}_{j} \Phi_{j},
\end{equation}

\noindent where $\{\sigma_{j}\}$ is a set of real-valued weights that depend on the finite-difference operators under consideration.

The semi-discrete system we want to solve at all points except at the interface is then
\begin{eqnarray}
\partial_{t} \Phi_{j} &=& \Pi_{j}, \nonumber \\
\label{second_ODE}\partial_{t} \Pi_{j} &=& (D^{2}\Phi)_{j} \;\;\;\;\;\; j = 1 \ldots N-1,
\end{eqnarray}
where $D$ is any finite-difference operator that approximates the derivative operator to some order $q\ge1$  satisfying the summation by parts
property (SBP from now on) \cite{Kreiss-Scherer:1977,Strand199447,gko1995}. That is, it satisfies the discrete counterpart of the integration by parts property
\[
<\Psi,D\Phi> + <D \Psi,\Phi> :=  \Psi_{N}\Phi_{N} - \Psi_{0}\Phi_{0}.
\]
Alternatively, instead of $D^2$, we could use a second-order operator $D_2$ approximating the second derivative, which satisfies the corresponding SBP property, i.e. that can be written as in \cite{Mattsson_Nordstrom}
\begin{equation}
D_2=H^{-1}\left( -D^THD+RS\right),
\label{D2}
\end{equation}
 being $H=dx\; diag(\sigma_0,...,\sigma_N)$, $D$  an operator that approximates the first derivative, $R=diag(-1,0,...,0,1)$  and $S$ an operator that approximates the first derivative at the interface. This guarantees that the analogue of the integration by parts property for the second derivative holds, i.e. that $D_2$ satisfies
\begin{equation}
\label{SBPoperator}
<\Psi,D_2\Phi> = \Phi_{N}(D\Psi)_{N} - \Phi_{0}(D\Psi)_{0} - <D \Psi,D\Phi>.
\end{equation}
This is a better choice, since the operators have a smaller stencil and preserve the solution phase more accurately. In our scheme, we shall use the narrow diagonal SBP second-order operators obtained in \cite{Mattsson-Parisi}. 

If one could prove that the linear ODE system (\ref{second_ODE}) has eigenvalues with no positive real part and a complete set of
eigenvectors, then one could apply any discrete-time integrator, and as a result obtain a stable numerical evolution
for the whole system. For a more detailed description of the theory see, for instance, \cite{gko1995}.
A way to check those conditions is to find a energy norm that is either constant or decreases in time. This is the procedure we shall use to implement our scheme.

Using (\ref{l2_norm}),  the discretized version of the energy norm (\ref{Energy}) becomes
$$
\mathcal{E}=<\Pi,\Pi>+<D\Phi,D\Phi>
$$
and we get its time derivative as
\begin{eqnarray}
\partial_t\mathcal{E}&=& <\partial_{t}\Pi,\Pi>+<D\partial_{t}\Phi,D\Phi>\nonumber\\
&=&<D_2\Phi,\Pi>+<D\Pi,D\Phi>\nonumber\\
&=&{\Pi}_{N} (D\Phi)_{N} - {\Pi}_{0} (D\Phi)_{0}.
\label{energy0}
\end{eqnarray}
Here we have used  equations (\ref{second_ODE}) and (\ref{SBPoperator}). In order to preserve this norm during evolution we need to cancel these terms, since the contributions in the RHS of this equation come from each side of the interface.

In contrast to this first-order hyperbolic and to the parabolic case, it does not seem possible for second-order systems to
control the energy by introducing on each side only terms proportional to the difference of the fields and their normal
derivatives at each interface without modifying the standard energy as in Mattsson  et al. \cite{Mattsson-Ham}.
Thus, following \cite{reula2011}  we introduce our first modification by adding terms at the interface as follows:

\[
\partial_{t} \Pi_{j} =  (D_{2}\Phi)_{j} + \frac{1}{dx \, \sigma_{0}}\delta_{j0} (D\Phi)_{0}  - \frac{1}{dx \, \sigma_{N}}\delta_{jN} (D\Phi)_{N},
\]
where $\delta_{ij}$ is the Kronecker delta. With this modification the interface terms are cancelled in equation (\ref{energy0}) and remaining constant  the energy norm, but they introduce no interaction between the two sides of the interface, and so the solution we get would just bounce back at the interface (the energy is conserved and if one interface point can not possibly influence the corresponding point at the other side, the pulse has to bounce back). However, eliminating the interface term means that we can now concentrate on adding terms that, while preserving/decreasing the energy norm, introduce an interaction at the touching points of the grid in such a way that the wave can pass through the interface.
We must therefore introduce a term that couples the two sides, namely a penalty term that forces the values at both extremes to coincide.
The simplest one that satisfies this requirement is
\begin{eqnarray}
\label{corrected_Pi}\partial_{t} \Pi_{j} &=& (D_{2}\Phi)_{j} + \frac{1}{dx \, \sigma_{0}}\delta_{j0} (D\Phi)_{0}  -  \frac{1}{dx \, \sigma_{N}}\delta_{jN} (D\Phi)_{N} \nonumber \\
&&                             - L(\Pi_{0}-\Pi_{N})(\frac{1}{dx \, \sigma_{0}}\delta_{j0}-   \frac{1}{dx \, \sigma_{N}}\delta_{jN}),
\end{eqnarray}
\noindent where $L$, which we call the {\sl interaction factor}, is a positive real constant to be chosen as large as possible in order to make the interaction as strong as possible. In this way we penalize the difference on both sides of the interface and drive them to coincide through a very large exponential factor while keeping the energy bounded, as follows
\begin{eqnarray}
\label{energy1}
\partial_{t}\mathcal{E}&=&<\partial_{t}\Pi,\Pi>+<D\partial_{t}\Phi,D\Phi>\nonumber\\
&=&<D_2 \Phi,\Pi>+<D\Pi,D\Phi>+ {\Pi}_{0} (D\Phi)_{0}-{\Pi}_{N} (D\Phi)_{N} - L \,(\Pi_0-\Pi_N)^2\nonumber\\
&=& -L \,(\Pi_0-\Pi_N)^2.
\end{eqnarray}

The limitation of how large  $L$ can be chosen comes from the fact that a too large negative real part would make the system unstable by making a large contribution to the eigenvalues along the negative real axis, thus making explicit time integration schemes fall outside their stability region, or making the needed time step prohibitively small. For explicit schemes, the value of $L$ should not be larger than  $L=1$, since it contributes to the CFL factor
as much as the principal part.  This value turns out not to be large enough for our scheme, giving unacceptably large errors in the form of bounces at the interface for a resolution that describes appropriately the solution. Thus we use larger factors and resort to a semi-implicit method which would free us from the CFL limitation.

Summarizing, the system of ordinary differential equations described in (\ref{second_ODE}) with the proposed correction introduced in (\ref{corrected_Pi}),
 is evolved using a third-order IMEX method \cite{Ascher1997151,pareschi}, specifically the one called IMEX-SSP3(4,3,3) L-stable scheme  as per
\cite{pareschi}. For the spatial discretization we propose finite-difference operators approximating the second
derivative obtained in \cite{Mattsson-Parisi}.

In order to confirm the correctness of our approach we compare the results of the method proposed with the corresponding traditional explicit third order Runge-Kutta
method verifying the presence of the bounces at the interface with the choice of $L=1$. In addition, for the spatial discretization,
we  compare the  choice of the spatial operator that approximates the second derivative
 with the use of first derivative operators \cite{Lehner:2005bz,Strand199447,Optimized} applied twice.
We do this because in some systems where off-diagonal terms occur in the Laplacian, or lower-order terms are present,
one might want to use a single operator for every derivative.

 Furthermore we compare the results achieved applying these methods with  the traditional evolution obtained using the standard $FO$-{\sl scheme}
\cite{Carpenter1999341},
which consists of adding penalty terms to the dicretization of equation (\ref{FO_system}) at the interface points, causing the energy to be preserved, namely

\begin{eqnarray}
\partial_{t} \Phi_j &=& \Pi_j, \nonumber \\
\partial_{t} \Psi_j &=& D\Pi_j -\frac{1}{2}\frac{\delta_{j0}}{dx \, \sigma_0}\left((\Pi_0-\Pi_N) + (\Psi_0-\Psi_N)\right) \nonumber \\
&& +\: \frac{1}{2}\frac{\delta_{jN}}{dx \, \sigma_N}\left((\Pi_N-\Pi_0) - (\Psi_N-\Psi_0)\right),\nonumber\\
\partial_{t} \Pi_j  &=& D\Psi_j-\frac{1}{2}\frac{\delta_{j0}}{dx \, \sigma_0}\left((\Pi_0-\Pi_N) + (\Psi_0-\Psi_N)\right) \nonumber \\
&& \label{FO_SAT}-\: \frac{1}{2}\frac{\delta_{jN}}{dx \, \sigma_N}\left((\Pi_N-\Pi_0) - (\Psi_N-\Psi_0)\right).
\end{eqnarray}

Finally, as  mentioned in the Introduction, Mattsson et al. developed a finite difference second order method that includes a SAT technique to treat the interface and the boundary conditions (hereafter $SO$-{\sl  Mattsson et al.'s scheme}).

To be able to compare both schemes,  we use the notation that we introduced in this section  for summarizing the $SO$-{\sl  Mattsson et al.'s scheme}. We consider a domain that contains an interface at $x=0$ and
we denote by $\Phi^{(1)}$ and $\Phi^{(2)}$ the numerical solutions at the left and right grids respectively. Hence, the
conditions that must be imposed on the interface are given by:
\begin{eqnarray}
 I_1&\equiv &\Phi_N^{(1)}-\Phi_0^{(2)}=0,\nonumber\\
 I_2&\equiv &\Pi^{(1)}_N-\Pi^{(2)}_0=0, \label{IF}\\
 I_3&\equiv &(RS\Phi^{(1)})_N+(RS\Phi^{(2)})_0=0, \nonumber
 \end{eqnarray}
\noindent where $R$ and $S$ are  defined   in  equation (\ref{D2}). In addition, at the left and right boundaries we should have:
 \begin{equation}
 (RS\Phi^{(1)})_0=0\qquad \mbox{and} \qquad(RS\Phi^{(2)})_N=0.
 \label{boundary}
 \end{equation}
The semi-discretization of the differential equation,  imposing (\ref{IF}) and (\ref{boundary}) becomes\footnote{Notice that some signs in (\ref{eqdif}) are different from those appearing in \cite{Mattsson-Ham}, which contain some typos.}:
\begin{eqnarray}
 \Phi_{tt}^{(1)}&=&D_2\Phi^{(1)}+H^{-1}\big (\tau {\bf e}_N I_1+\beta (RS)^T{\bf e}_N I_1+\gamma {\bf e}_N  I_3+\sigma {\bf e}_N
I_2\big ) \nonumber\\
&=& D_2\Phi^{(1)}+H^{-1}\big (\tau {\bf e}_N I_1+\beta S^T {\bf e}_N I_1+\gamma {\bf e}_N  I_3+\sigma {\bf e}_N
I_2\big ), \nonumber\\
 \Phi_{tt}^{(2)}&=&D_2\Phi^{(2)}-H^{-1}\big (\tau {\bf e}_0 I_1+\beta (RS)^T{\bf e}_0 I_1-\gamma {\bf e}_0  I_3+\sigma{\bf e}_0
I_2\big )\nonumber\\
&=&D_2\Phi^{(2)}-H^{-1}\big (\tau {\bf e}_0 I_1-\beta S^T {\bf e}_0 I_1-\gamma {\bf e}_0  I_3+\sigma{\bf e}_0
I_2\big ).
 \label{eqdif}
\end{eqnarray}
\noindent Note that the term containing $S^T$ introduces the penalties in the evolution of the interior points on both sides of the grids.


Using (\ref{eqdif}), we calculate the energy proposed by Mattsson et al. and  verify the correctness of the stability conditions derived in
\cite{Mattsson-Ham},
namely
\begin{eqnarray}
 \gamma &=& -\frac{1}{2}= -\beta,  \nonumber \\
 \tau &\leq& -\frac{1}{2 \, \alpha_M \,dx }, \nonumber \\
 \sigma &\leq& 0 \, , \nonumber
\end{eqnarray}
\noindent where the positive constant $\alpha_M$ is given by  Lemma 2.3 in \cite{Mattsson-Ham}. As we compare the schemes, it arises that $\alpha_M$ should
satisfy $\alpha_M \leq \sigma_0$.

We report on the findings in the following section.



\section{Tests}
\label{tests}


We test the method by running simulations both in one and two dimensions ($1D$ respectively $2D$). For the $1D$ case all the runs are performed on a circle of length 2 (i.e. the domain is the interval $[0,2]$, where the last grid point is identified with the first one). For the $2D$ simulations  the domain is a torus and the grid consisted of a $2\times2$ square with the $x=0$ face identified with the $x=2$ face, and similarly for the $y$ coordinate. In this case, for simplicity, one of the interfaces, namely the one corresponding to the $y$ direction, is treated using penalties, while for the $x$ direction we use periodic operators.
In all the runs
the number of points and the order of the finite-difference operators employed guarantee a good enough resolution for cases where the solution has a high frequency\footnote{Here we aim at an accuracy of about one part in $10^{3}$ for $10$ periods. Enough to keep the phase without appreciable error for about $10$ crossing times.}.


\subsection{Initial-data sets}


For the purpose of analyzing convergence it is sufficient to choose {\bf smooth} initial data. We therefore choose the following data
\begin{itemize}
\item \textbf{1D smooth initial data}
\[
\Phi_0(x) := 4^{12} \, x^{12}(x-1)^{12},
\]
\[
\Pi_0(x) :=  \partial_{x} \Phi_0(x).
\]
\end{itemize}

On the other hand, for the purpose of comparing realistic situations and in order to analyze how the method keeps the phase of the solution, we take the following {\bf rough} and highly variable data:
\begin{itemize}
\item \textbf{1D rough initial data}
\[
\Phi_0(x) := e^{{- 8^{2} (x-0.5)^{2}}} \cos(50 \pi x) ,
\]
\[
\Pi_0(x) :=  \partial_{x} \Phi_0(x) ,
\]
corresponding to a rough pulse propagating to the left.

\item \textbf{2D rough initial data}
\[
\Phi_0(x,y) :=  e^{{- 8^{2} ((x-1.5)^{2}+(y-1.5)^{2})}} \cos(50 \pi (y-x)),
\]
\[
\Pi_0(x,y) :=  \frac{1}{\sqrt{2}}(\partial_{x} \Phi_0(x,y) - \partial_{y} \Phi_0(x,y)).
\]
\end{itemize}


\subsection{Space discretizations}


As mentioned in the Introduction, in this section we perform runs for  different schemes and  different choices of the space discretization.
 We compare the results to
\begin{itemize}
 \item The traditional $FO$-{\sl scheme}.
 \item The new second-order formulation  presented in this paper (hereafter $SO$-{\sl interface approximation}) with a second derivative operator $D_2$.
 \item The $SO$-{\sl interface approximation}  where the second derivative is approximated by the first derivative operator $D$ applied twice
(i.e. $D^2$) with and  without dissipation. Here we use the same discretization as in $FO$-{\sl scheme}.
\item The $SO$-{\sl  Mattsson et al.'s scheme} with a second derivative operator $D_2$.
\end{itemize}

In all the runs we use a very accurate finite-difference operator, in particular, the first derivative operator is an optimized
operator of order eight in the interior and order four at points in the boundary \cite{Lehner:2005bz,Strand199447,Optimized}. The second derivative operator used is of order eight in the interior and order six at
the boundary \cite{Mattsson-Parisi}; this operator comes also with a first-order companion that is used for the boundary
contributions, both of these satisfy SBP with the same norm. In the comparison with $SO$-{\sl  Mattsson et al.'s scheme} we use the finite difference operator of 6th-order accuracy in the interior and 5th-order accuracy at the boundary (see \cite{Mattsson-Ham})

Furthermore, the choice of these operators was made in order to preserve
the correct phase of the solution on long-time runs, and to be able to test the contribution to the error coming form the
interaction term, with the smallest possible interference from the contribution to the error of the space  discretization of the derivatives.


\subsection{Time integration}


As noted in section 2, we use two time integrators, a traditional  Runge-Kutta  scheme and an IMEX one.
The necessity of an IMEX scheme comes from the fact that the interaction factor has to be very large, hence stiff, in order to achieve good accuracy.

\begin{figure}[htbp!]
\begin{center}
\includegraphics[width=3.5in]{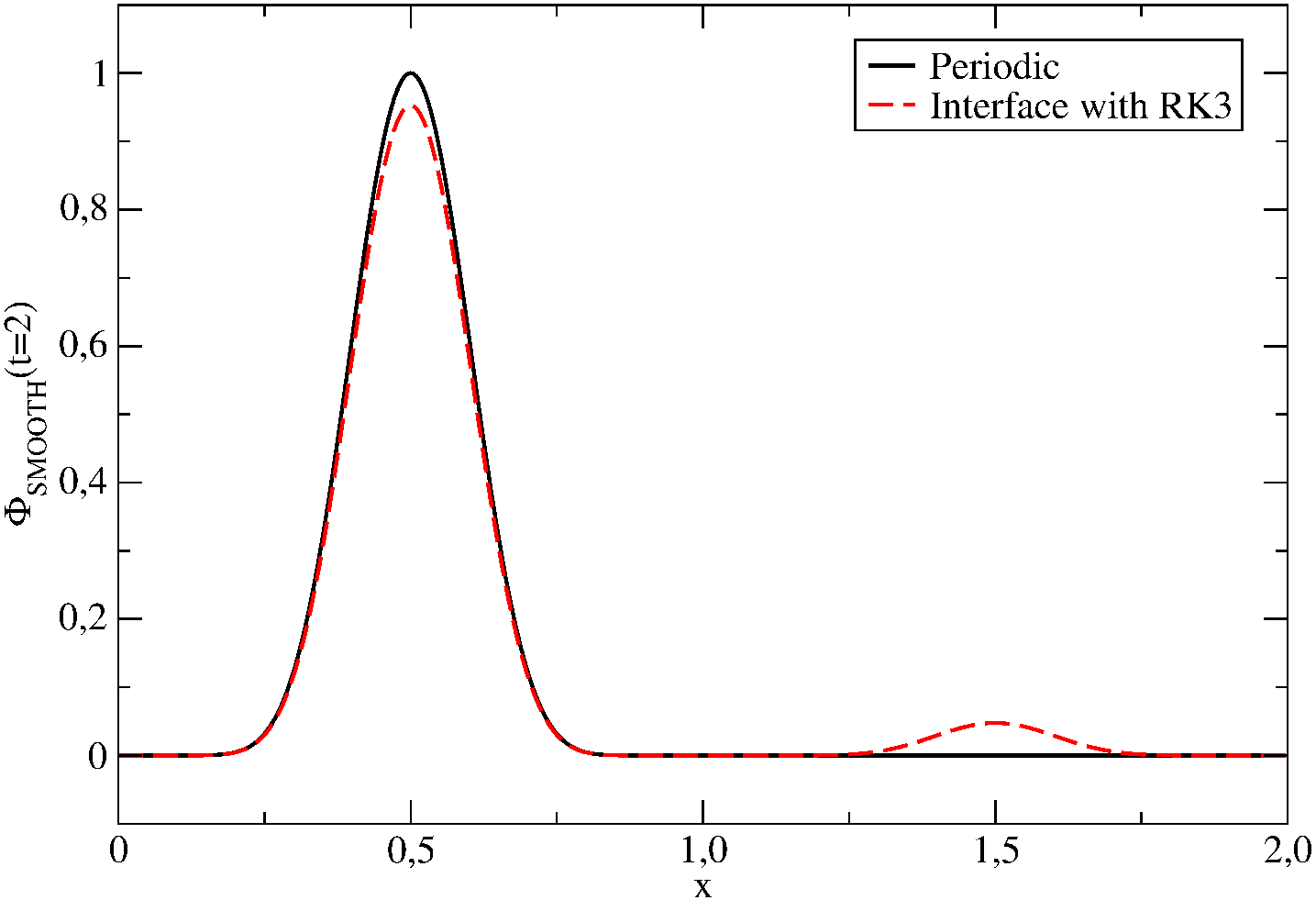}
\caption{{\bf Comparison of periodic and $SO$-{\sl interface approximation}  runs using the traditional third-order Runge-Kutta method whit
$L=10$.}}
\label{Comparison-RK}
\end{center}
\end{figure}

To visualize this,  we implement the $SO$-{\sl interface approximation}  with  an interaction factor of $L = 1$ and $L = 10$,  and we evolve equation (\ref{corrected_Pi}) using a  smooth initial data and a traditional third-order Runge-Kutta. The runs  were performed with a resolution of $N=640$ points and with $dt=2.5 \times 10^{-5}$ (CFL $=0.008$). In the plot below, Figure \ref{Comparison-RK}, we show both the periodic solution, i.e. the exact solution of the wave equation (\ref{wave}), and the $SO$-{\sl interface approximation}. The extra bump to the right is the bounce of a fraction of the solution at the interface.
\vspace*{0.7cm}

It is possible to reduce the error to a very small amount by enlarging the interaction factor, but at the expense of losing
efficiency, since for the traditional Runge-Kutta scheme the time step needed for stability becomes significantly smaller. In fact,
we observe that  the errors  fall to very small values for an interaction factor a thousand times larger if we use, in the traditional Runge Kutta, a time step a thousand times smaller.

\begin{figure}[htbp!]
\begin{center}
\includegraphics[width=3.5in]{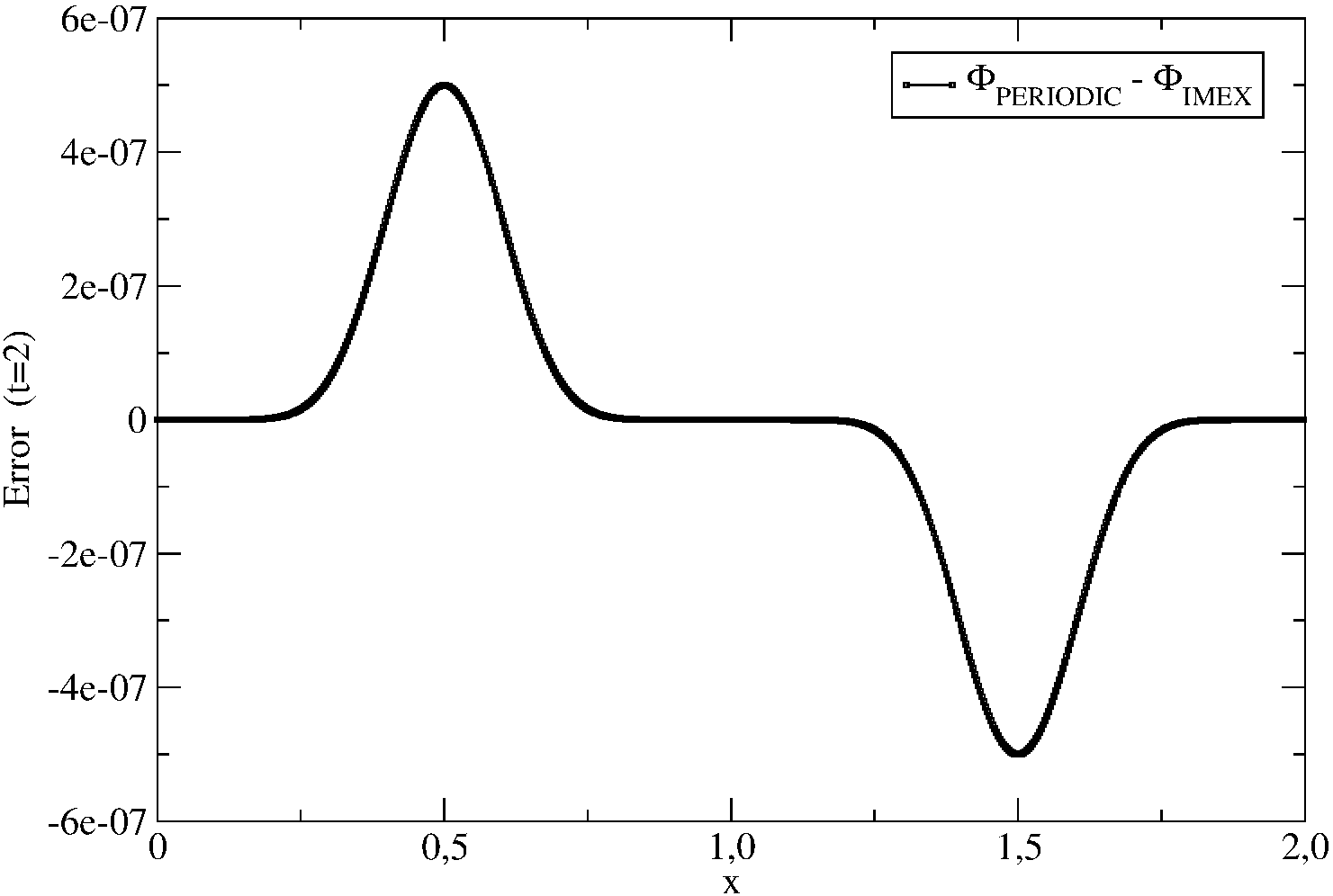}
\caption{{\bf Error of the interface run (i.e., difference with the periodic run), using IMEX-SSP3(4,3,3) L-stable time integrator with an
interaction factor $L=10^6$.}}
\label{ComparisonIMEX}
\end{center}
\end{figure}

Thus to avoid small time steps, while allowing larger interaction factors, semi-implicit methods are needed.
For this reason, we use in our implementation a method among those called IMEX, \cite{Ascher1997151,pareschi}, specifically, the one called
IMEX-SSP3(4,3,3) L-stable scheme  presented in \cite{pareschi}. This method permits us to explicitly solve stiff parts of the equations while keeping the other terms as usual in traditional Runge-Kutta schemes. In the plot below, Figure \ref{ComparisonIMEX}, we show the error of the interface (i.e. the difference with the periodic run) using the IMEX method.

From now on, for all runs using $SO$-{\sl interface approximation},  we present the
results evolved with the IMEX-SSP3(4,3,3) L-stable time integrator, while all runs for the $FO$-{\sl scheme},
  all the periodic runs and  comparison with $SO$-{\sl Mattsson et al.'s  scheme} ( in section 3.8) are performed  with the traditional third or fourth-order Runge-Kutta integrator. Note that the only
term that needs to be treated implicitly with this IMEX method is the term proportional to $L$, that is, just the last boundary
term in equation (\ref{corrected_Pi}).

\subsection{Convergence}

In the absence of the interaction term, we expect the error to be of the form $e = f_{1}dt^{p} + f_{2}dx^{q}$, where $p$ depends on the time integrator used and $q$ on the space discretization of derivatives.

The convergence rate is calculated as
\begin{equation}
Q=\ln\left(\frac{\|\Phi^{(h)}-\Phi^{(h/2)}\|_{l^2}}{\|\Phi^{(h/2)}-\Phi^{(h/4)}\|_{l^2}}\right)/\ln(2),
\end{equation}
\noindent where $\Phi^{(h_i)}$ is the numerical solution with grid spacing $h_i$.

In our case, we expect $p\geq 3$ for the IMEX algorithm. The precise value depends on the nature of the solution, in particular the size of the solution near the boundary (where the implicit part of the algorithm is used) in comparison with the size of the solution in the interior of the grid. Since for  the space discratization (in the case of  $D^2$) we use a fourth-order accurate operator at the boundary and eighth-order accurate in the interior, $q\geq 5$.

For stability reasons, the CFL condition on the explicit integrator is such that we need to scale $dt$ as $dx$, so we expect a convergence index of the order of
three. Alternatively, we might fix a sufficiently small $dt$ and increase the space resolution, which allows us to study in an independent way space convergence. In this case we would expect a convergence index of the order of five.
Any smaller convergence factor must result from the interface treatment. For most of the convergence tests we used very smooth initial data, since the $f_{1}$ and $f_{2}$ functions depend on high derivatives of the exact solution.

We start by analyzing the convergence of the method for the $1D$ case with runs of $640, 1280$ and $2560$  points using the
smooth data thus in   $FO$-{\sl scheme} as $SO$-{\sl interface approximation} with $D^2$ and $D_2$ operator, and with CFL and $dt$ fixes.

\begin{figure}[htbp]
\begin{center}
\includegraphics[width=2.9in]{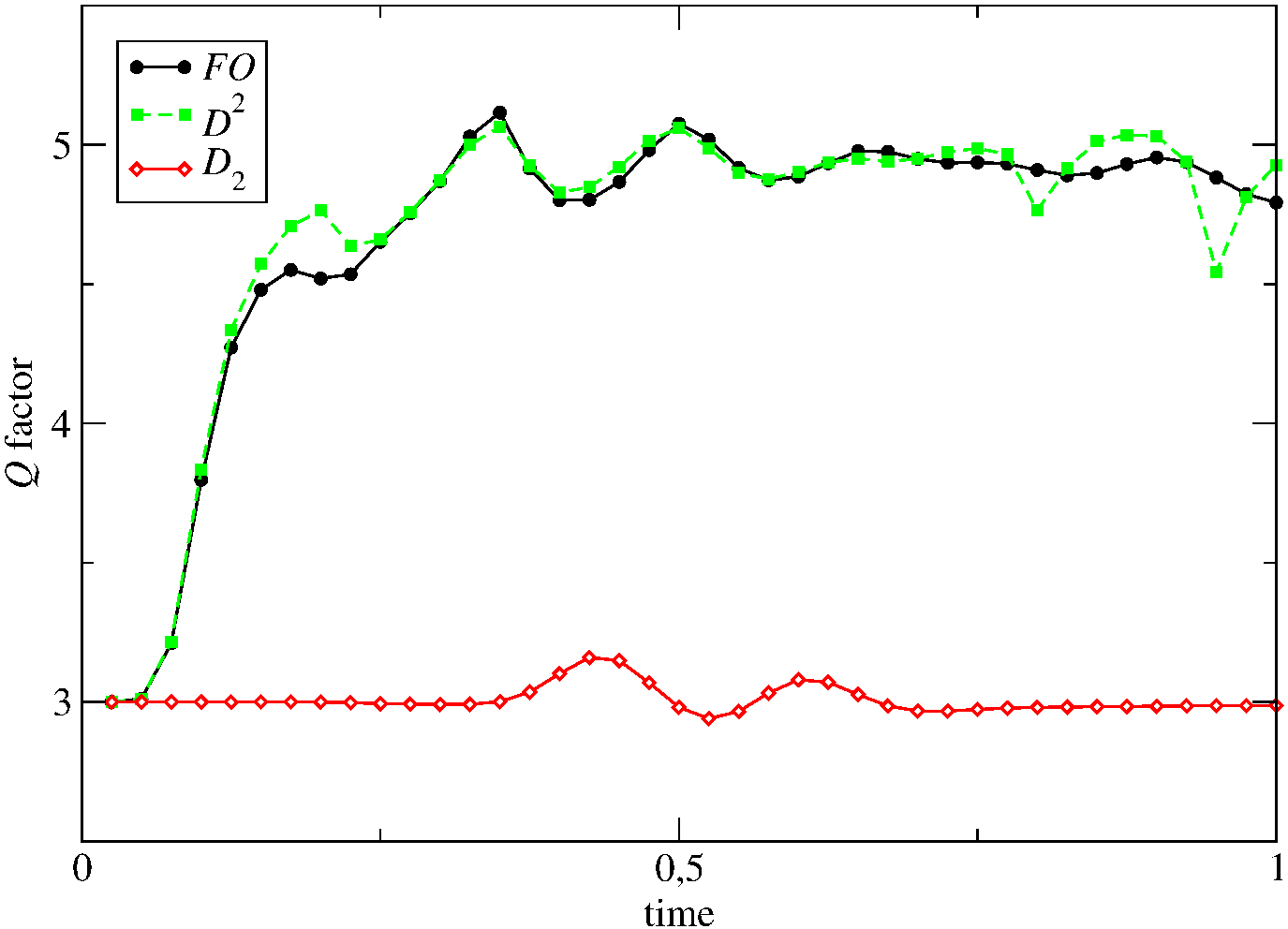}{(a)}
\vspace{0.2in}
\includegraphics[width=2.9in]{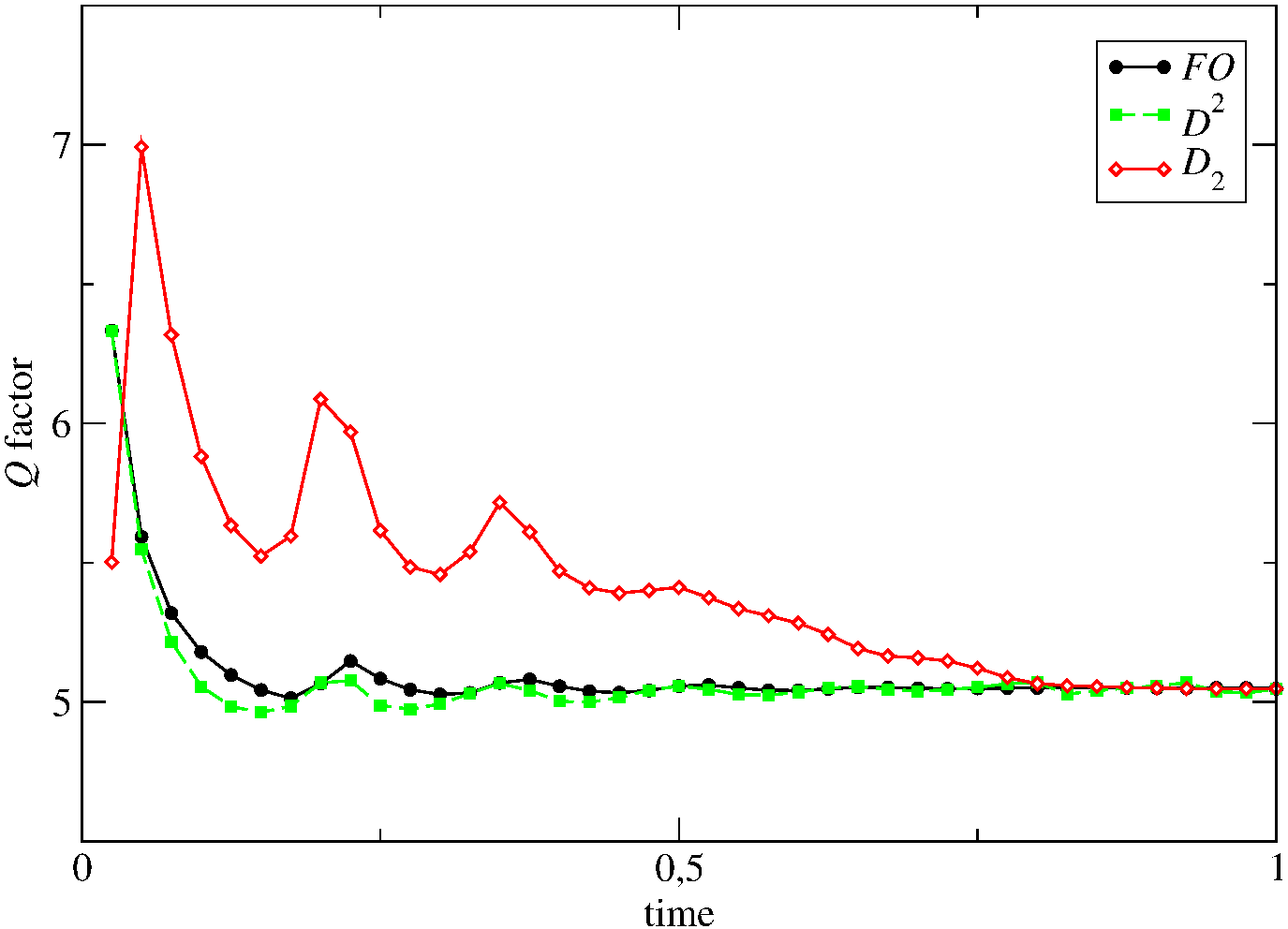}{(b)}
\caption{{\bf Comparative plot of the convergence factor for the $1D$ system, for: first-order formulation $FO$-{\sl scheme}, $SO$-{\sl interface approximation} using $D^2$ and $D_2$ respectively, at (a) CFL, and (b) $dt$ fixes.}}
\label{ConvergenceIMEX}
\end{center}
\end{figure}

From  Figure \ref{ConvergenceIMEX} we see that, for $FO$-{\sl scheme} and  $SO$-{\sl interface approximation} with $D^2$ operator
keeping CFL constant (0.08), the convergence factor starts at a value of $3$ while the pulse is in the interior, meaning that the
main contribution to the error comes from time discretization.  By the time the solution reaches the boundary the $Q$ factor
climbs to $\sim 5$, which means that there the space discretization is the primary contribution to the error. For $SO$-{\sl interface approximation} with $D_2$
second-order case, however, the convergence remains constant around $3$, implying that during the whole run the derivative
operator the  contribution to the error is negligible.

On the other hand, for fixed $dt=2.5 \times 10^{-5}$, we observe that during the whole run the error is dominated by the space operators, and the convergence factor starts at a high value, close to $8$, corresponding to the time when the pulse has not reached the boundary; and falling to $5$ when it does.

For the $2D$ case we performed runs of $640\times 640$, $1280\times 1280$ and $2560\times 2560$ points with the rough data.

\begin{figure}[htbp]
\begin{center}
\includegraphics[width=3.3in,angle=0]{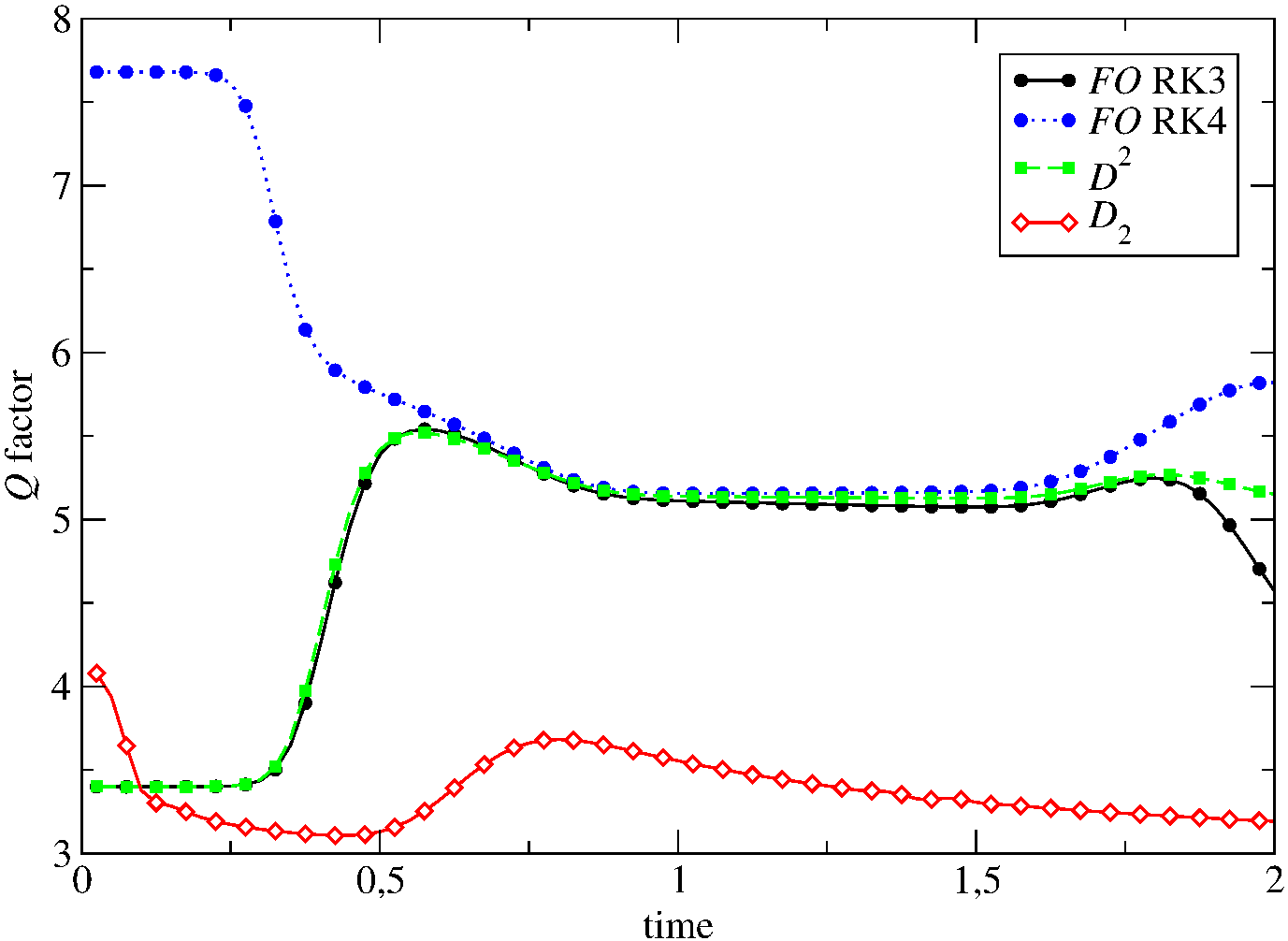}
\caption{{\bf Comparative plot of the convergence factor for the $2D$ system, at fixed CFL $=0.08$. Plotted are the first-order formulation $FO$-{\sl scheme} and $SO$-{\sl interface approximation} using $D^2$ and  $D_2$ respectively.}}
\label{Convergence2D}
\end{center}
\end{figure}

From Figure \ref{Convergence2D}, we observe very similar behavior as in the $1D$ case. Before the wave reaches the boundary, for all discretizations using the third-order Runge-Kutta integrators, both conventional or IMEX,  convergence is dominated by the time discretization with a $Q$ factor close to $3$, climbing to $\sim 5$ as the pulse reaches the interface (where the space discretization is the one contributing the most to the error).

We also performed a run, for comparison, using a fourth-order Runge-Kutta method for the first-order system. In this case we observe that in the interior the convergence  improves and starts close to $8$. Here the time integrator is more accurate and hence the space discretization becomes more important. As the pulse reaches the boundary, we again obtain a $Q$ factor of $5$.

Convergence alone is not enough to guarantee that we are approaching the correct solution.
That is, in principle, the limit of our finite-difference scheme does not need to coincide with the continuum equation (because of
the boundary terms which grow with resolution). Thus, it is necessary to analyze convergence against the true solution, which we
do in the next section.


\subsection{Accuracy}
\label{accuracy}


Here we compare methods for realistic data, namely the rough initial data given above, both for the $1D$ and $2D$ cases.

For the $1D$ case we evolve the solution up to $t=2.0$, at which point the solution has moved to the left and the pulse has completely passed the interface located at $x=0$.
In the $2D$ case a pulse is sent in an oblique direction to the interface to check whether the scheme preserves the correct phase in this case and does not introduce, for instance, an
excess bounce.

For comparison we performed a run using periodic boundary conditions with eighth-order centered-difference operators with
$N=5120$ points, or $5120\times5120$ for the $2D$ case (referred to as $P_{5120}$ in both cases). This is used as the reference
solution against which we compare all the other runs. For these last simulations, interface conditions are used with $N=640$,
$1280$ and $2560$ points for both the $FO$-{\sl scheme} system and the $SO$-{\sl interface approximation} (denoted by $FO_{640}$, $FO_{1280}$, $FO_{2560}$
and $D_{2\; 640}$, $D_{2\; 1280}$, $D_{2\; 2560}$ respectively). In addition, for the $SO$-{\sl interface approximation}, we perform
simulations using both $D^2$, and $D_2$ operators. All the runs are performed
with an interaction factor $L=10^6$ and keeping the CFL factor constant ($0.08$).

\begin{figure}[htbp!]
\begin{center}
\includegraphics[width=3.5in]{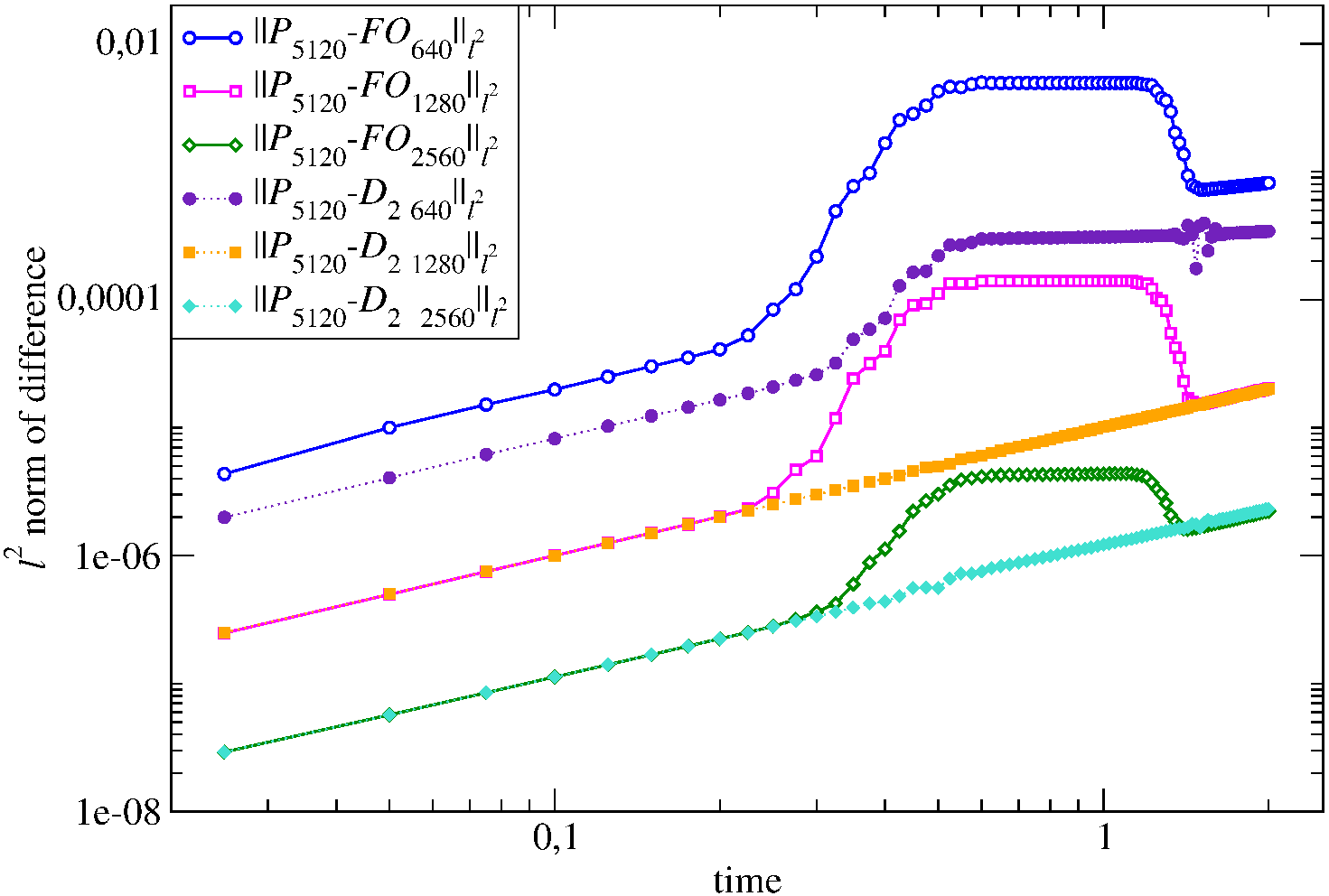}
\caption{{\bf $l^{2}$-norm of the error for several different $1D$ runs. }}
\label{L2_comparison_IMEX_long}
\end{center}
\end{figure}

In Figure \ref{L2_comparison_IMEX_long}, we show a comparison of the $l^{2}$-norm of the error for the two different cases under consideration: the standard $FO$-{\sl scheme} with a third-order Runge-Kutta time integrator, and $SO$-{\sl interface approximation} with the $D_2$ operator that uses the IMEX-SSP3(4,3,3) L-stable scheme. Note that before the pulse has reached the interface, the two methods are comparable, but as soon as the wave reaches and passes through the boundary, the solution obtained using our second-order method improves the accuracy by at least one order of magnitude. This shows that the interface treatment proposed here competes very well with the traditional $FO$-{\sl scheme}.

\begin{figure}[htbp!]
\begin{center}
\includegraphics[width=3.5in]{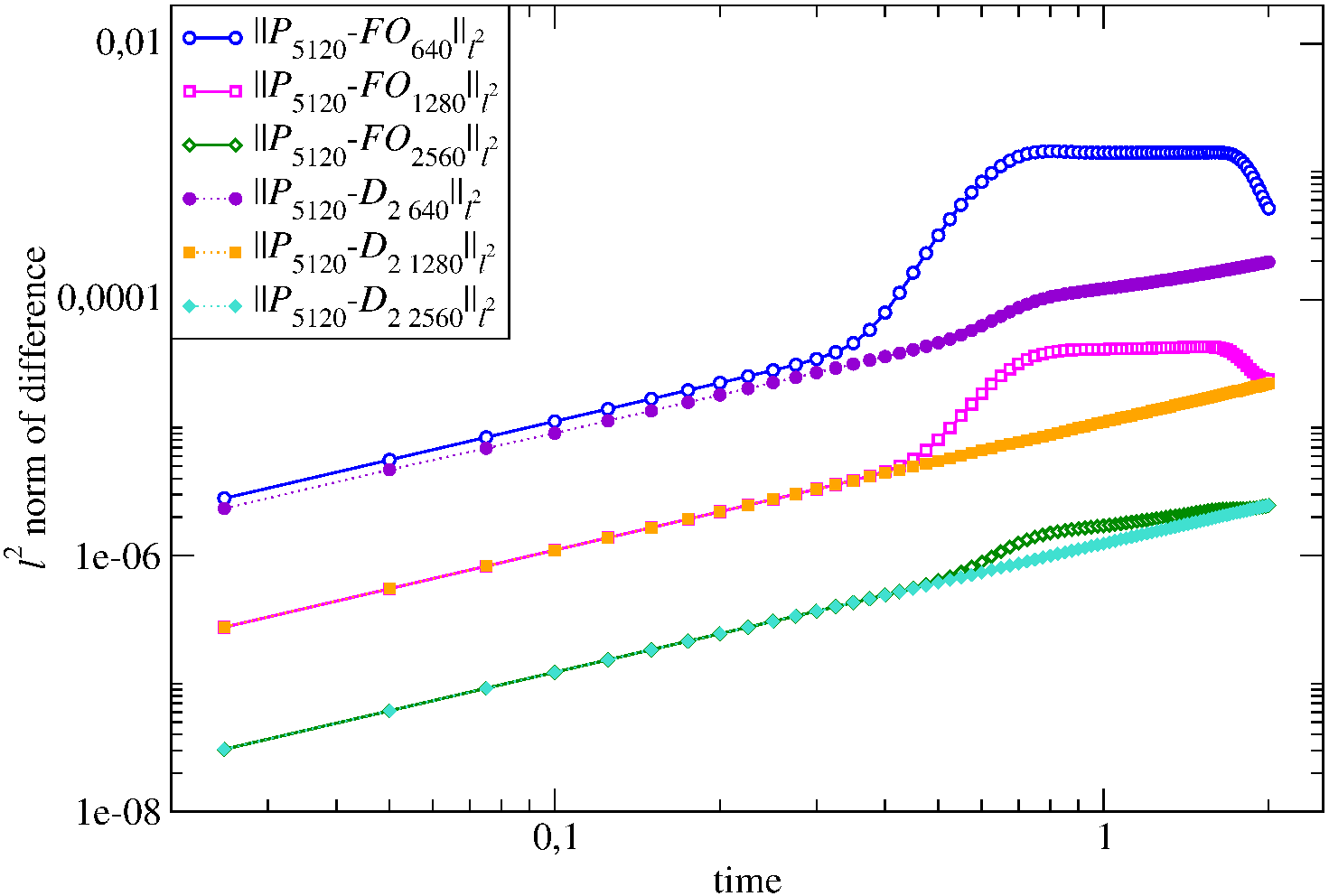}
\caption{{\bf $l^{2}$-norm of the error for several different $2D$ runs.}}
\label{L2_comparison_2D}
\end{center}
\end{figure}

We observe the same behavior for the $2D$ case, displayed in Figure \ref{L2_comparison_2D}. Again our method and the $FO$-{\sl scheme} behave similarly in the interior region, but ours is superior to the SAT after the pulse passes the interface.

\vspace{-0.1cm}

\subsection{Energy decay}


The present scheme is energy-diminishing at the semi-discrete approximation level. This implies that if a stable time integrator is used with a sufficiently small time step the energy given by (\ref{Energy}) should decrease only at a rate given by the penalty term, plus, perhaps noticeable, the inherent dissipation of the time integrator.
So here we study such a decay, showing that it is indeed very small, as one would infer from the method's accuracy.

\begin{figure}[htbp!]
\vspace{.8cm}
\begin{center}
\includegraphics[width=3.5in]{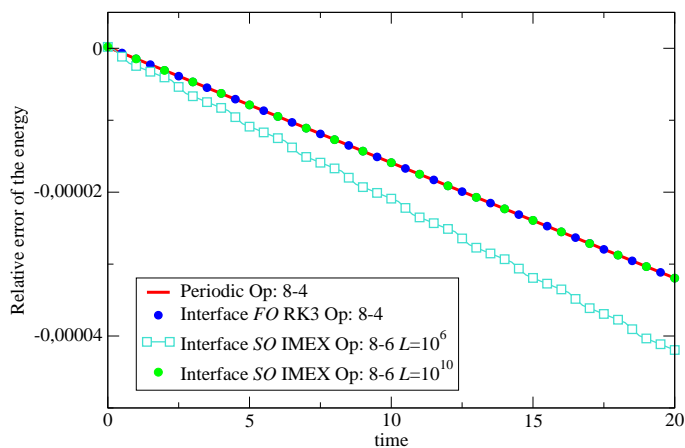}
\caption{{\bf Relative error of the energy decay for the rough initial data for four different scenarios: periodic boundary conditions, traditional  $FO$-{\sl scheme}, and $SO$-{\sl interface approximation} using $D_2$ for two choices of the interaction factor $L$.}}
\label{L2_Norm_IMEX}
\end{center}
\end{figure}
Figure \ref{L2_Norm_IMEX} shows the behavior of the relative error of the energy, i.e.
$(E-E_{initial})/E_{initial},$
on longer runs: ten times the previous ones.

These runs were performed with fixed CFL $=0.08$, and a resolution of $5120$ points. As
expected, the decay is very small, and it improves considerably for larger values of the interaction factor. For a value
$L=10^6$, the energy decays at a faster rate than with the first-order SAT scheme, which
coincides with the decay given by a periodic treatment. However, if we increase the value of $L$, the decay approaches the
periodic one, and if we take $L=10^{10}$ the three decays (periodic solution, $FO$-{\sl scheme}, and $SO$-{\sl interface approximation}) are
indistinguishable. So most of the decay is due to the inherent-Runge-Kutta integrators, and both the standard third-order and the
IMEX one seem to have the same dissipation.

\begin{figure}[htbp!]
\vspace{0.3cm}
\begin{center}
\includegraphics[width=3.5in]{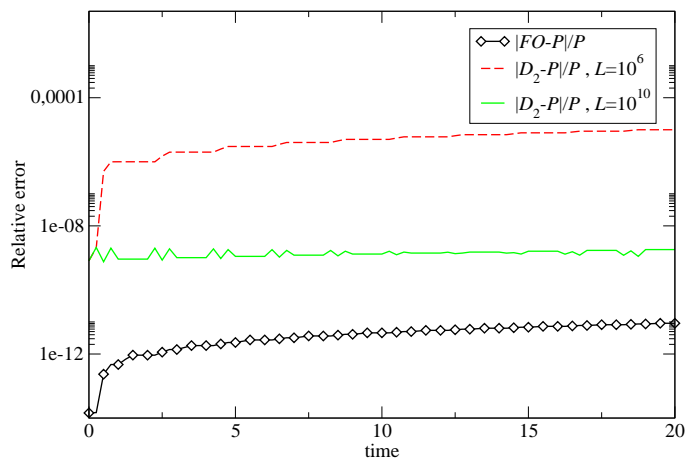}
\caption{{\bf Energy relative error (compared to the periodic-solution energy) for the rough initial data for three different scenarios: $FO$-{\sl scheme} and $SO$-{\sl interface approximation} using $D_2$ for two choices of the interaction factor $L$.}}
\label{energy_per_rough}
\end{center}
\end{figure}

Finally, in order to account only for the decay associated to the method, we show in Figure \ref{energy_per_rough} the relative error of the energy compared to that of the periodic solution, i.e.
$(E-E_P)/E_{P}$

Here we see that the $FO$-{\sl scheme} is the one that best approximates the periodic energy, while $SO$-{\sl interface approximation} deviates from it. This difference decreases, however, if we take a larger $L$, showing, once more, that the larger the interaction factor, the better the proposed method fits the solution.

\subsection{Dissipation}
\label{Dissipation}


It is worth noticing that, for all the runs performed so far, it was not necessary to introduce any artificial dissipation, for we have been considering a linear problem with constant coefficients and smooth data, and therefore there was no noise introduced by high frequency modes.
However  if we were dealing with a nonlinear equation or one with non-constant
coefficients, we might find high frequency oscillations around the correct solution. It is well known that adding dissipation to $D^2$ we achieve the same result than $D_2$.
As an example of this, we used the $D^2$
scheme instead of the $D_2$ used above, with the rough data.

\begin{figure}[htbp!]
\begin{center}
\includegraphics[width=3.5in]{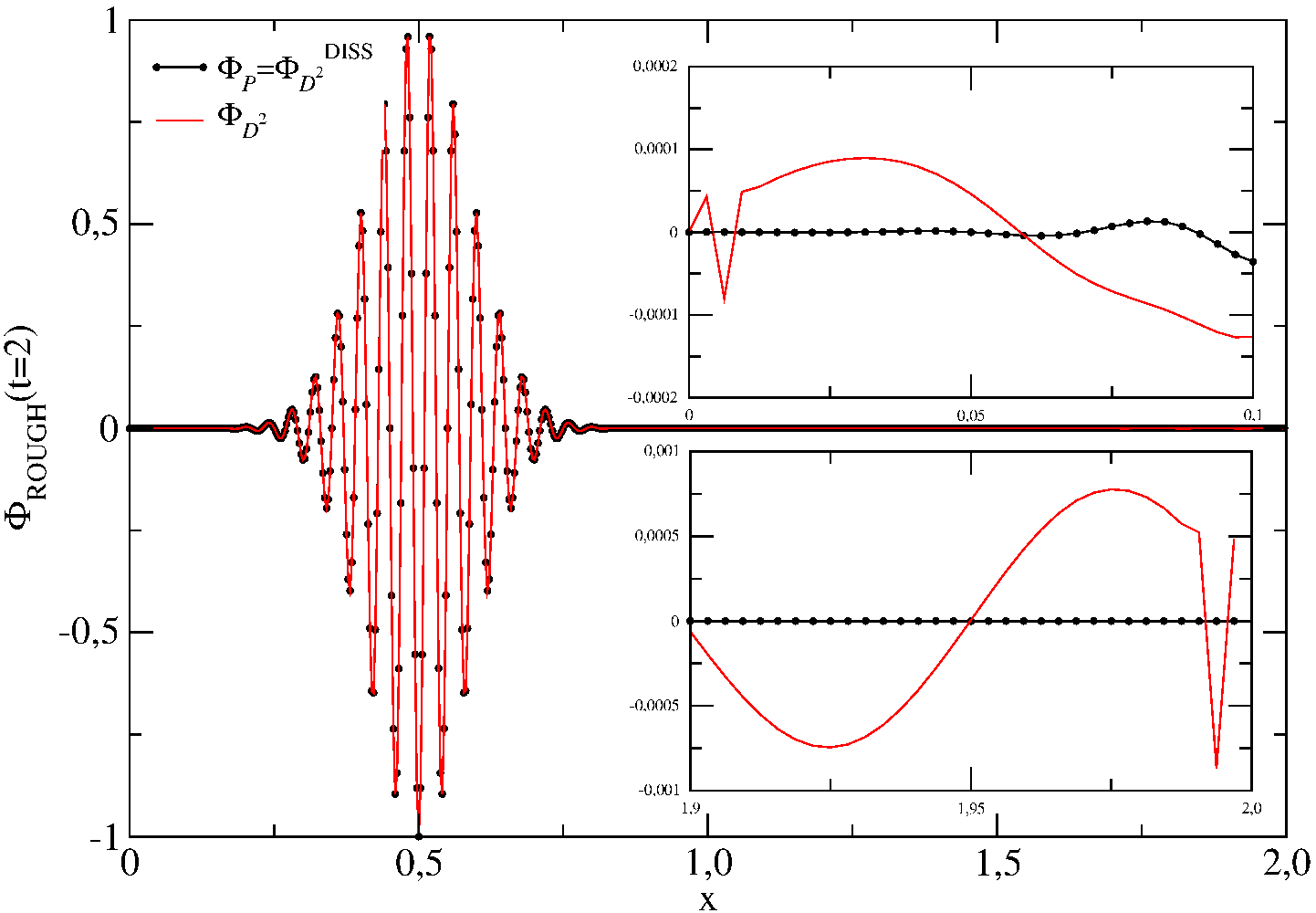}
\caption{{\bf Comparison of the solution at $t=2$ using $D^2$ for the spatial operator with and without dissipation, with the periodic one.}}
\label{Comparacion_DD_DD_DISS}
\end{center}
\end{figure}

We see in Figure \ref{Comparacion_DD_DD_DISS} that using this
operator introduces some numerical noise to the solution, diminishing its quality. We note that the solutions without dissipation
are almost indistinguishable except near the interface, where we include a zoomed sector to the right in  Figure \ref{Comparacion_DD_DD_DISS}  to show the disagreement. As known,  this noise is removed by using Kreiss-Oliger  dissipation \cite{gko1995}, that is, by adding to the equations a term proportional
to a large power of the Laplacian operator. This term contains a factor that depends on the resolution so that the error produced
is of the same or smaller order as the rest of the terms in the approximation.

In particular, we used the one that corresponds to the accuracy we are using for the finite-difference operators \cite{KO_diss,Lehner:2005bz,Optimized}, namely,  eighth-derivative dissipation $Q_d = -\sigma_d dx^{9} \Delta^{4}$, where $\Delta$ is a finite-difference operator that approximates the Laplacian to first-order accuracy. The runs used for this comparison were performed with a resolution of $640$ points at CFL fix ($0.08$), and $\sigma_d=100$.

\begin{figure}[htbp!]
\begin{center}
\includegraphics[width=3.5in]{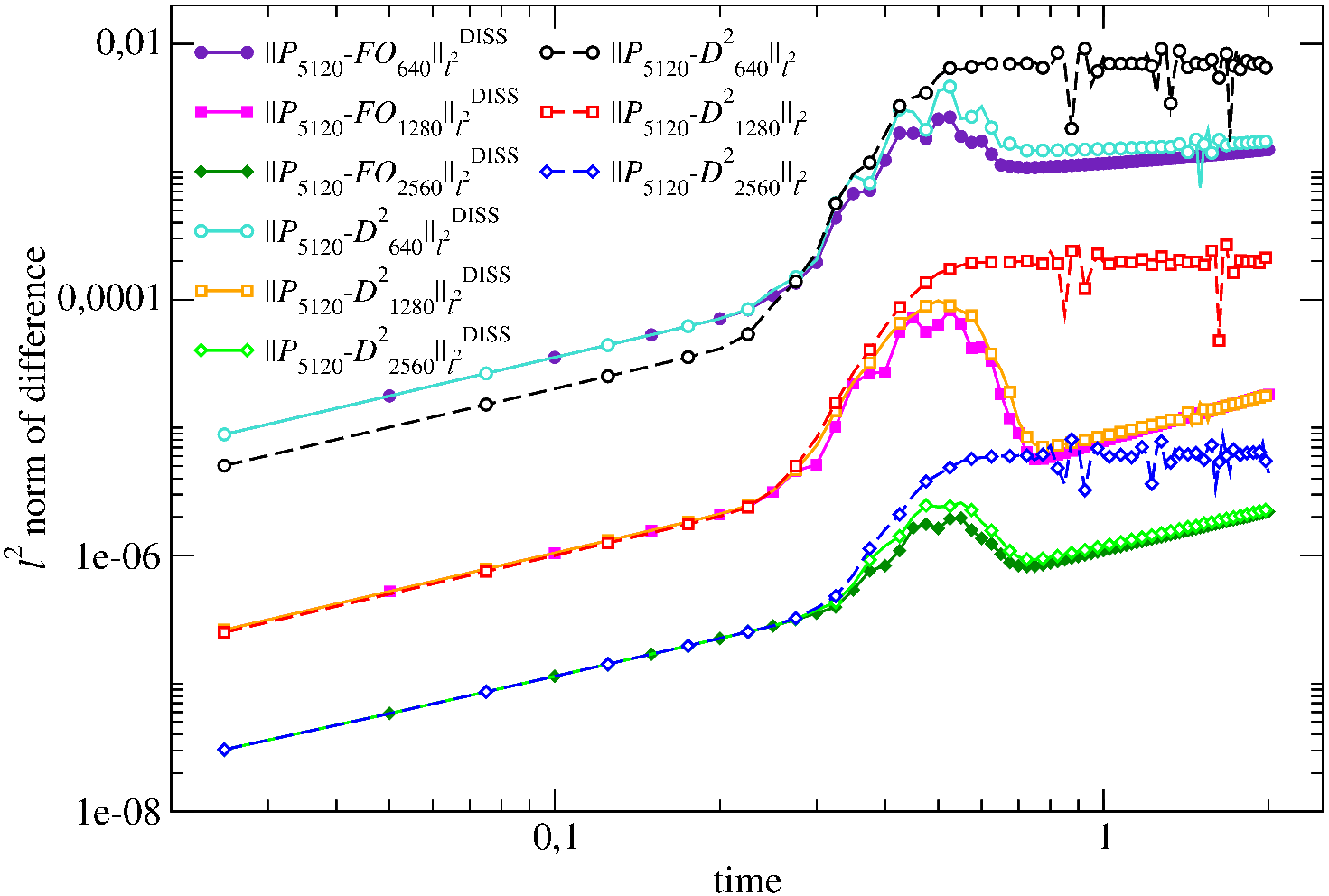}
\caption{{\bf Behavior of the relative error of the $l^{2}$-norm for runs with operator $D^2$ with and without dissipation.}}
\label{L2_Norm_DIS}
\end{center}
\end{figure}

Figure \ref{L2_Norm_DIS} below shows the $l^2$-norm of the error using $D^2$ with and without dissipation, as well as the error for the
$FO$-{\sl scheme} with dissipation.

We calculated the error by comparing a periodic run with $5120$ points against interface runs with
$640$, $1280$ and $2560$ points. These were done keeping the CFL factor fix ($0.08$) and using an interaction factor
$L= 10^6$. We see that adding the dissipative term improves the accuracy by one or two orders of
magnitude. The errors calculated with dissipation for both $FO$-{\sl scheme} and $SO$-{\sl interface approximation} systems are almost the same.

Also, by comparing with
Figure \ref{L2_comparison_2D}, we see that the errors for the $SO$-{\sl interface approximation} with $D^2$ and $FO$-{\sl scheme} with a dissipation are similar to the error
calculated $SO$-{\sl interface approximation} with the second-order operator $D_2$. Thus, as expected, methods that use dissipation are competitive with the $D_{2}$
discretization.


\subsection{Comparison with \citeauthor{Mattsson-Ham}'s method}

When implementing the  $SO$-{\sl  Mattsson et al.'s scheme}, we choose $\sigma=0$
$\alpha_M=\sigma_0$ and $\tau=-\frac{6.0}{dx\sigma_0}$ and in our $SO$-{\sl  interface approximation} we choose an interaction factor of $L=10^{10}$.
 These runs are performed with a fixed factor CFL $= 0.08.$

Furthermore we  use the  smooth initial data described in section 3.1  and  the finite difference operators   of second-order accuracy  and, of 6th-order  accuracy in the interior and 5th-order at the boundary,  calculated by Mattsson et al. in \cite{Mattsson-Ham}. For our scheme we use  the  operators of second-order accuracy and, of 8th-order accuracy in the interior and 6th-order at the boundary with a first-order companion, calculated in \cite{Mattsson-Parisi}.

In Figures \ref{dif_and_error}.a and \ref{dif_and_error}.b we show the error of both method when comparing the numerical solutions obtained using the $SO$-{\sl  Mattsson et al.'s scheme} ($\Phi_{M}$) and our one ($\Phi_{D_2}$) with the corresponding periodic solutions of   same difference operator's accuracy ($P_{2560}$)  for a resolution of $2560$ grid points.
 
  We see that both methods are very similar with a better performance  in our scheme when  the pulse reaches the interface (see Figure \ref{dif_and_error}.b). Since both methods basically only differ in the treatment of the interface,  the difference becomes appreciable  in that  region.
The method developed by Mattsson et al. treats the interface using penalty techniques on $\Phi$, $\Pi$ and $D\Phi$. These are present  not only in the evolution of the terms of the interface, but also in the evolution of the interior point, being the  amount of points  subjected to  the operator accuracy,
whilst our method we only needs to communicate the $\Pi$ field, even though the IMEX scheme requires that this information is passed twice.

\vspace{.15in}
\begin{figure}[htbp!]
\begin{center}
\includegraphics[width=3.5in]{matt-flor-orden2}{(a)}
\nonumber
\end{center}
\end{figure}

\begin{figure}[htbp!]

\begin{center}
\includegraphics[width=3.5in]{matt-flor-orden6-8.eps}{(b)}
\caption{{\bf $l^2$-norm of the error corresponding to $SO$-{\sl  Mattsson et al.'s scheme} and to  $SO$-{\sl  interface approximation}  for $N=2560$ with (a) a second-order accuracy difference operator for both scheme and (b)  a 6th and 8th-order accuracy difference operator  respectively}}
\label{dif_and_error}
\end{center}
\end{figure}


\section{Applications}
In this section, we present two applications of the method developed above, namely, a variable coefficient problem (gauge-wave) and a wave packet propagating on the surface of the 2-sphere.
\subsection{Gauge-wave}

As an application of our method involving variable coefficients, we consider a simple $1D$ test in numerical relativity: a linearized solution of Einstein's equations around a gauge-wave background with line element \cite{Tiglio-Lehner-Neilsen}
\begin{equation}
\label{gw_metric}ds^{2}=e^{A\,\sin(\pi(x-t))}(-dt^{2}+dx^{2})+dy^{2}+dz^{2}.
\end{equation}
This background describes flat spacetime, in which a coordinate transformation on the $(t,x)$ plane has been performed, with a sinusoidal dependence along  $t-x$. This gauge-wave problem provides us with a simple, yet non trivial, numerical test, for it is linear, the amplitude of the coefficients can be controlled by only adjusting the parameter $A$, and does not lead to any singularities. This test differs from those of the previous sections since in this case the coefficients depend both on space and time.

There are various papers that deal with this problem \cite{Lehner:2005bz,Tiglio-Lehner-Neilsen,CPST,Alcubierre}. Most of them use a method that involves a first-order formulation with periodic boundary conditions, except for \cite{Lehner:2005bz}, which uses a boundary treatment. A second-order scheme with boundary conditions for this gauge-wave problem was studied in \cite{Winicour0,Winicour1}. One aspect that these papers show is the exponential growth and loss of convergence displayed by the solution for large amplitudes.

In this section we will apply the method developed above to analyze this problem. We use the same approach as \cite{Lehner:2005bz} where perturbations of (\ref{gw_metric}) are considered and introduced in Einstein's equations in order to derive the linearized evolution equations for the fields. Here we study the short-time behavior since we are only interested in the stability of the method as the waves go through the interface.

The non trivial variables for this problem are the relevant components of the metric and its time derivative $(g_{xx},K_{xx})$, and the lapse $\alpha$. We consider, therefore, perturbations of the form
\begin{eqnarray}
g_{xx}&=&e^{A\,\sin(\pi (x-t))}+\delta g_{xx},\\
K_{xx}&=& \frac{A}{2}\,\cos(\pi (x-t))e^{\frac{A}{2}\,\cos(\pi (x-t))} + \delta K_{xx},\\
\alpha&=&e^{\frac{A}{2}\,\sin(\pi (x-t))} + \delta \alpha.
\end{eqnarray}

The resulting equations are
\begin{eqnarray}
\partial_{t} \Phi &=& \Pi + A\pi \,\cos(\pi(x-t)), \nonumber \\
\label{gw_2nd}\partial_{t} \Pi   &=& \frac{1}{\hat{\alpha}} \partial_x(\hat{\alpha} \partial_x \Phi)   \nonumber \\
&&- \frac{1}{2}\left( A\pi^{2}\,\sin(\pi(x-t)) + \frac{A^{2}\pi^{2}}{2}\,\cos^{2}(\pi(x-t))\right)\Phi \nonumber \\
&& -\frac{1}{2}A \pi \,\cos(\pi(x-t))\Pi,
\end{eqnarray}

\noindent where $\hat{\alpha}=e^{\frac{A}{2}\sin(\pi (x-t))}$, $\Phi=\delta \alpha/\hat{\alpha}$ and $\Pi=\delta K_{xx}$.

We perform several runs and compare the results for the $SO$ {\sl interface approximation} treated in this paper, using
first derivative operator applied twice ($D^2$) and the second derivative operator ($D_2$). The semidiscretization of the
second derivative term in (\ref{gw_2nd}) is thus of the form
\begin{equation}
\partial_x(\hat{\alpha} \partial_x \Phi) \approx D(\hat{\alpha} D\Phi),
\end{equation}
for the $D^2$ case.

For the $D_2$ case, on the other hand, we split the second derivative as
\begin{equation}
\partial_x(\hat{\alpha} \partial_x \Phi) \approx \hat{\alpha} D_2 \Phi + (\partial_x \hat{\alpha}) D_1 \Phi,
\end{equation}
where the derivative of $\hat{\alpha}$ is calculated analytically, and $D_1$ and $D_2$ are fully compatible finite-difference operators approximating the first and second derivative respectively ( see \cite{Mattsson-Parisi}). Note that, in certain circumstances,  the chain-rule form can experience instability but this is not the case,  thus for simplicity, we use the chain-rule to expand de $D_2$ operator despite there are narrow stencil formulations for this operator.

\begin{figure}[htbp!]
\begin{center}
\includegraphics[width=3.5in]{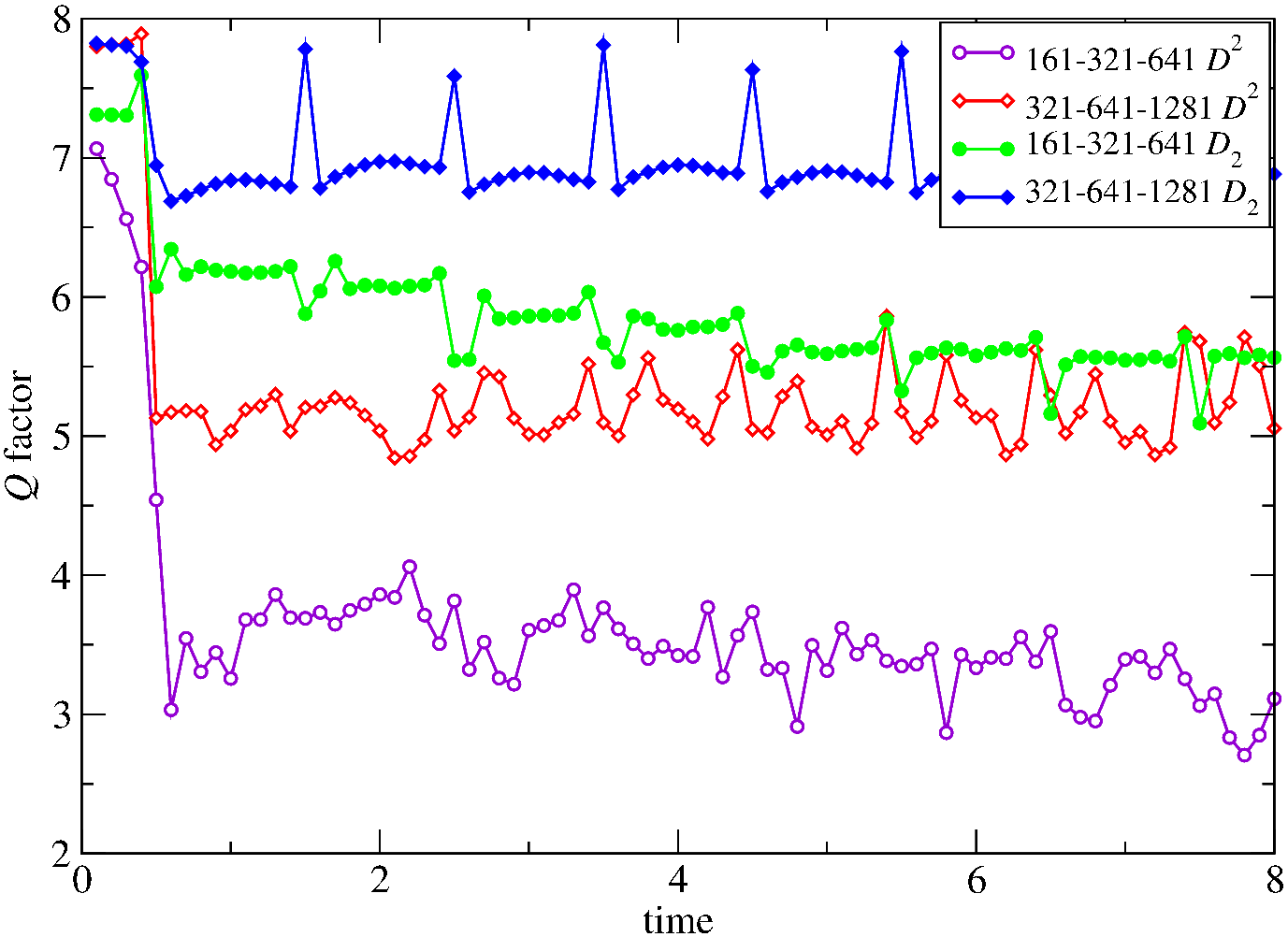}
\caption{{\bf Convergence factor for two different resolutions with  $D^2$ and $D_2$  in the $SO$-{\sl interface approximation} }}
\label{q_factor_gw}
\end{center}
\end{figure}

\begin{figure}[htbp!]
\begin{center}
\includegraphics[width=3.5in]{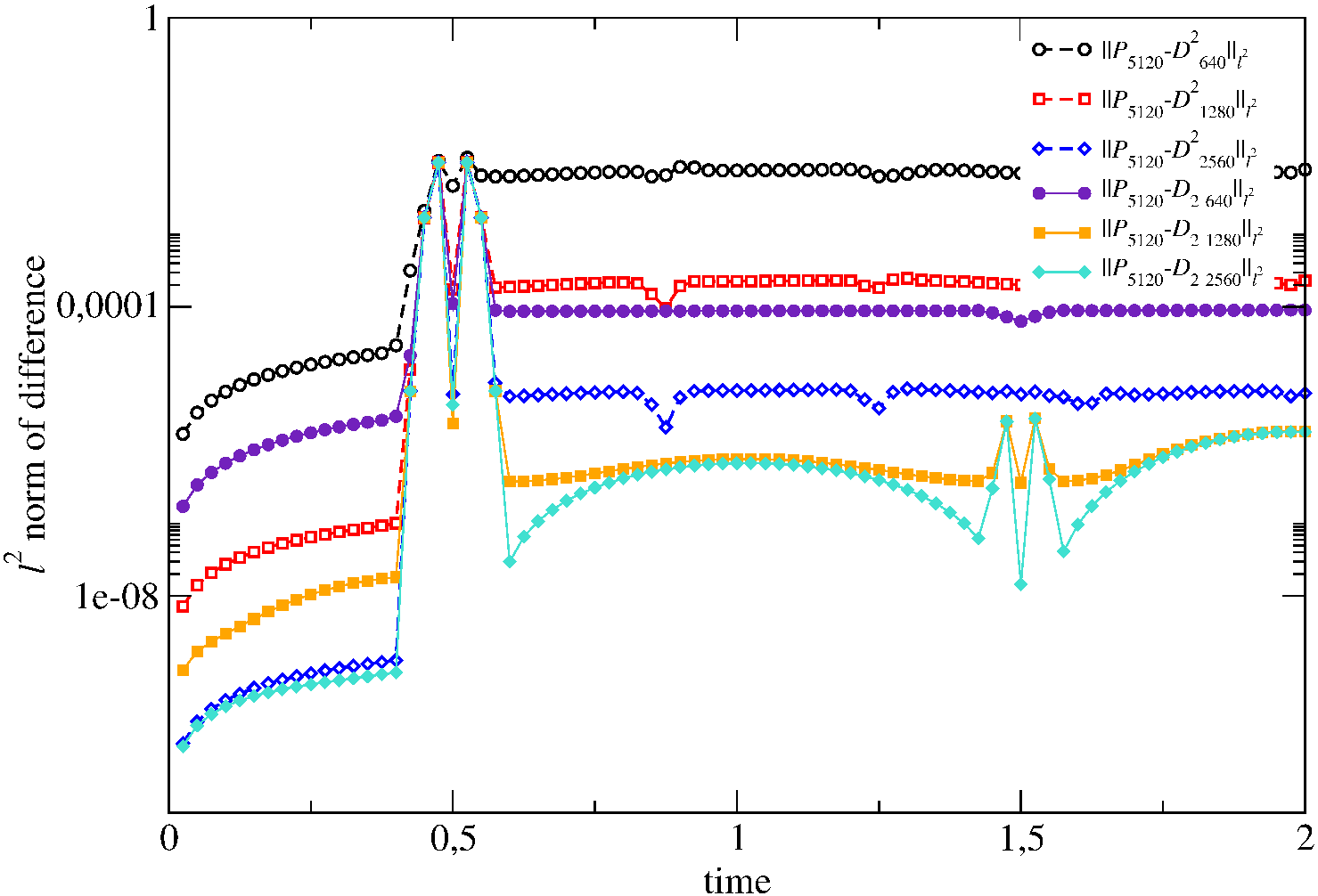}
\caption{{\bf $l^2$-norm of the error for various resolutions for the $SO$-{\sl interface approximation}.}}
\label{l2_norm_gw}
\end{center}
\end{figure}

We run the $SO$ {\sl interface approximation} using $N= 161,321$ and $641$ points and $N=321,641$ and $1281$ points both using $D^2$ and $D_2$ operators. In Figures (\ref{q_factor_gw}) and (\ref{l2_norm_gw}) we show the convergence factor and the $l^2$-norm of the error for these two cases compare with the periodic solution with $N=5120$ points, respectively. The runs were performed in the $[-1,1]$ interval, using  an amplitude of $0.5$, CFL $=0.01$ and the following smooth initial data

\begin{eqnarray}
\Phi_0(x) &:=& 100^{12} \, (x+0.6)^{12}(x+0.4)^{12},\\
\Pi_0(x) &:=&  \partial_{x} \Phi_0(x) .
\end{eqnarray}

For all the runs, except for the $D^2$ case with the lowest resolution, the convergence factors oscillate between the expected values of $8$ and $5$, consistent with the order of the difference operators used.
The $l^2$-norm of the error shows again that the $D_2$ formulation has a lower error compared to the $D^2$ case for all the resolutions considered.
This illustrates the superior performance of the present method even for variable coefficients, which opens a wide range of
possible applications.

\subsection{Waves in the $2$-sphere}

\begin{figure}[htbp!]
\begin{center}
\includegraphics[width=7.5cm]{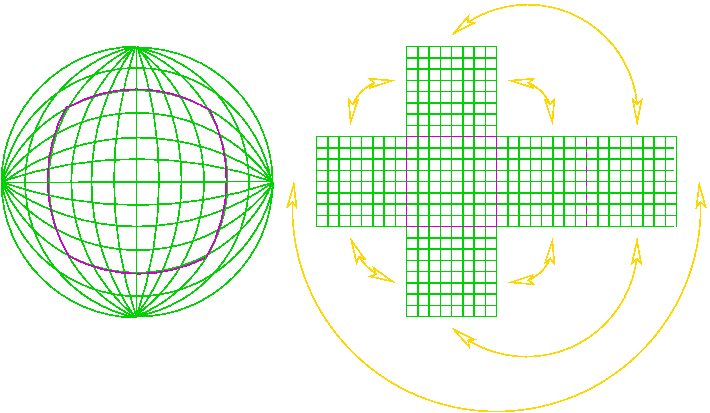}
\caption{{\bf Configuration of the grid patches used to cover the $2$-sphere.}}
\label{cubitos}
\end{center}
\end{figure}

As a second application of our method, consider the wave propagation on the surface of a $2$-dimensional sphere. We discretize the
domain using six identical square grids, one for each of the six patches whose faces are identified in such a way that they cover
the sphere (see Figure \ref{cubitos}). The idea is to evolve each patch
separately by independent processors and, by imposing the proper interface conditions, obtain the global evolution of the
solution.

\vspace{1cm}

\begin{figure}[htbp!]
\begin{center}
\includegraphics[width=4.5in]{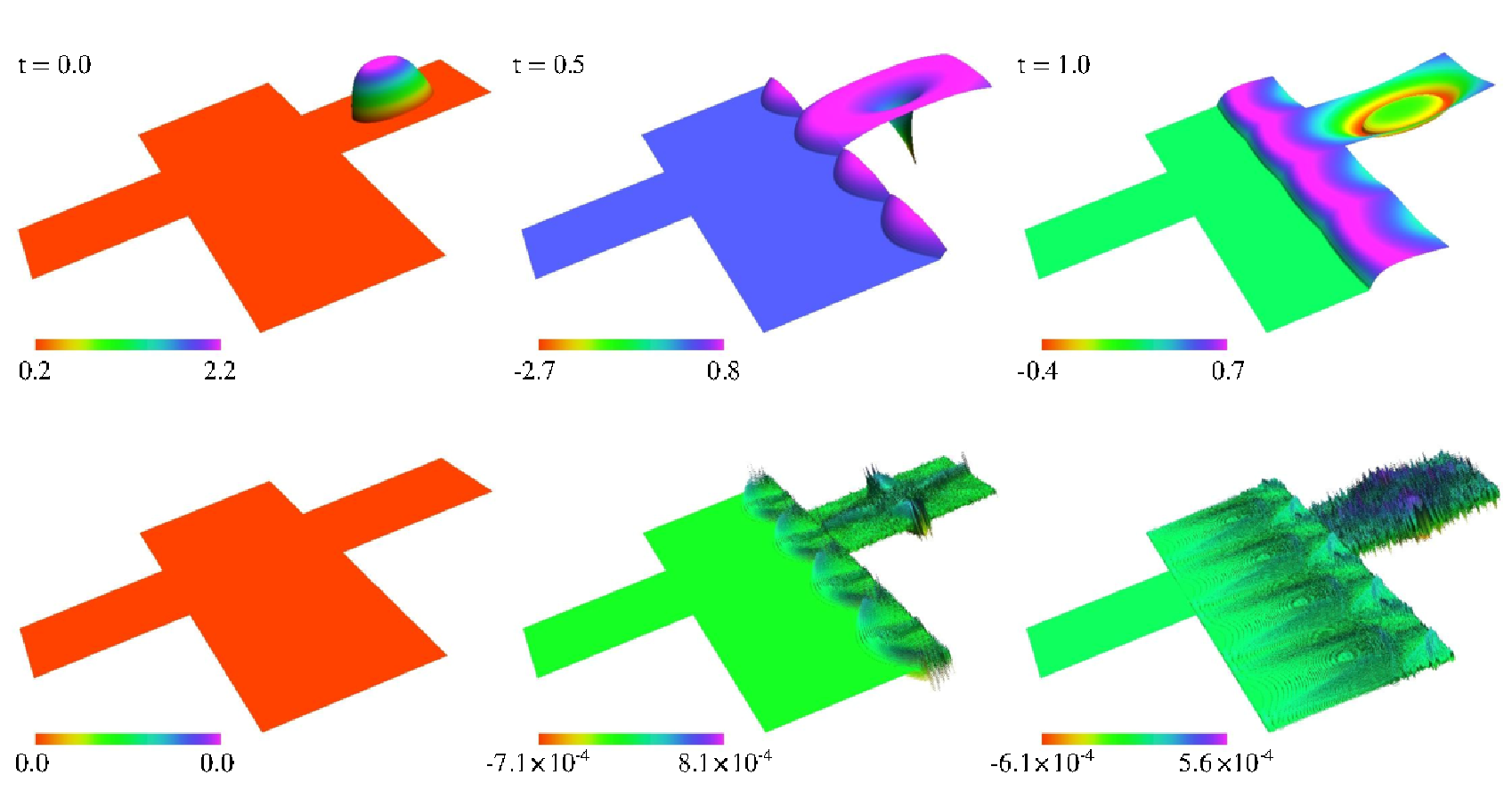}
\includegraphics[width=4.5in]{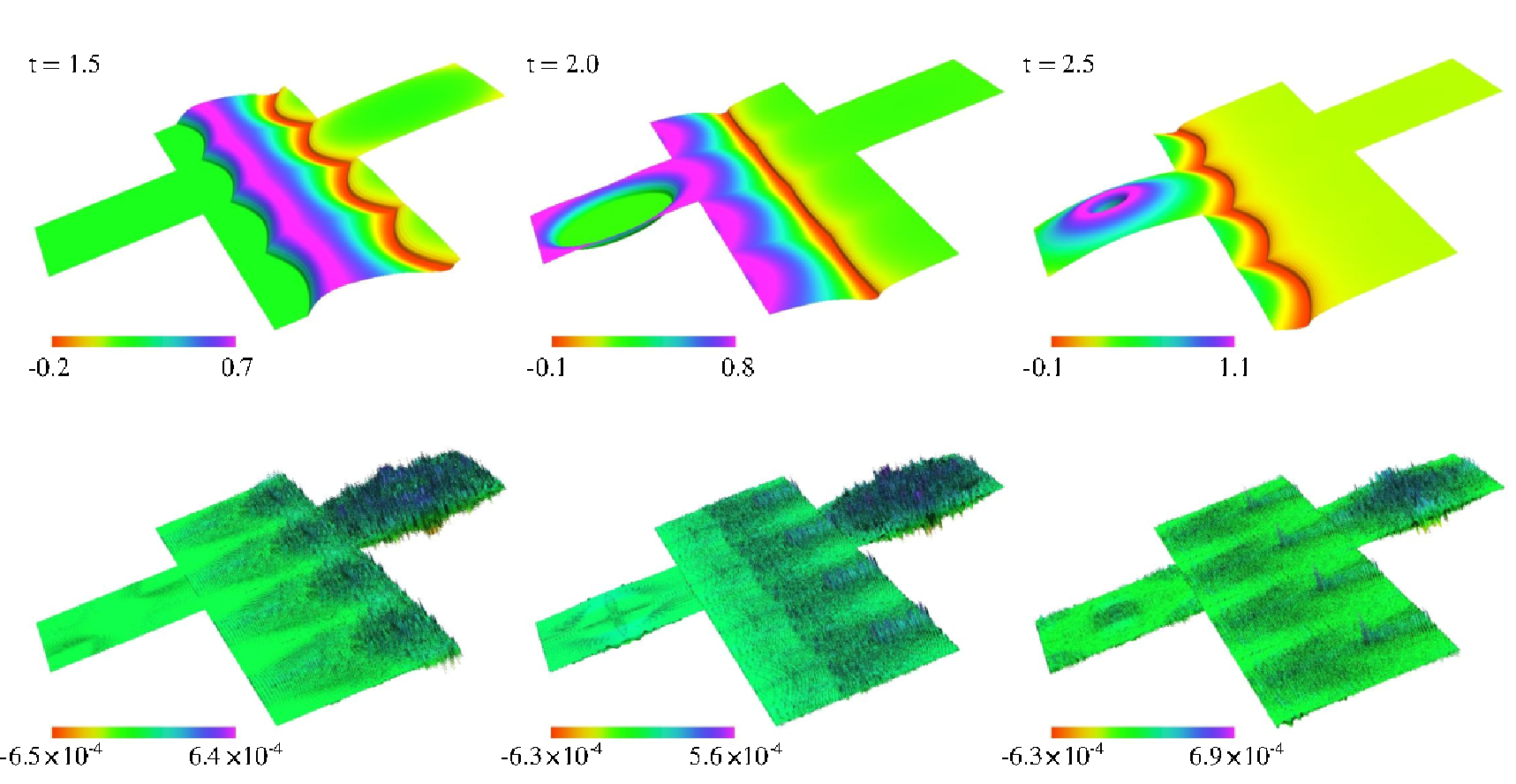}
\caption{{\bf Evolution of the wave packet in the surface of the $2$-sphere with our method (first and third row panels) and
the difference
between this solution and the one obtained with the SAT technique (second and fourth row panels)}.}
\label{wave_sphere}
\end{center}
\end{figure}

So, we consider the $2$-dimensional wave equation
\begin{equation}
\label{2d_wave} \partial^2_t u = \frac{1}{\sqrt{g}}\partial_a \left(\sqrt{g} g^{ab}\partial_b u \right),
\end{equation}
\noindent
with $g^{ab}$ the metric of the sphere in the coordinates of each patch and $g$ its determinant. We use the initial data on the
sphere given by

\begin{equation}
u(\theta,\phi,t=0) = \left\{
\begin{array}{l l}
  2\left((\theta^4 - k^4)/k^4 + 0.1\right) & \quad \mbox{if $\theta \leq k$}\\
  0.2 & \quad \mbox{if $\theta > k$ }, \end{array} \right.
\end{equation}

\noindent where $\theta$ and $\phi$ are the standard polar angles on the sphere, and $k = 0.5$, and apply the method introduced in
section (\ref{numericalscheme}) so that each interface is treated using the scheme presented in (\ref{corrected_Pi}). In Figure
\ref{wave_sphere} we show
the evolution of the solution for several different times (first and third row panels), as well as the difference
between this solution and the
one obtained treating the interface using the
standard SAT method \cite{Carpenter1999341} (second and fourth row panels). We see that the wave packet behaves nicely
as it passes through the
interfaces, and that the difference between the two methods
is of order $10^{-4}$. We also can see from the second and third row panels that the interface treatment in our method seems to be
as good as the
SAT scheme, since there is no appreciable difference in the interface,
the largest values of the error being the ones in the interior of the patch where the initial bump is located.

\vspace{1cm}


\section{Conclusions}


We have shown that it is possible to implement an interface scheme of  ``penalty'' type for the second-order wave equation similar to the ones
used for first-order hyperbolic and parabolic equations and for second-order wave equation with Mattsson et al.'method. Our scheme shares with them similar properties: Only data at
interface points need to be passed between grids, and convergence is ensured for linear, constant-coefficient
systems.

Our scheme was applied as well to a problem with non-constant coefficients (perturbations of a gauge-wave
background) and the wave propagation on the surface of a $2$-dimensional sphere.
The accuracy of the method seems to be as good as the accuracy of the finite-difference operators and of the time integrators used, and
competes favorably with both the usual $FO$-{\sl scheme} and $SO$-{\sl Mattsson et al.'s scheme} for all the cases we have tried.

Note that for the wave equation in both the $FO$-{\sl scheme} and $SO$-{\sl Mattsson et al.'s scheme} one must pass at the boundary many more quantities
than in
our scheme, namely, in addition to the time derivative fields, either all space derivatives or the normal derivative at the
boundary.
This fact is important for multi-block parallelizations in several dimensions, for it implies that one obtains the same solution
quality while sharing among different computational grids only a  small fraction of the data one would need for a comparable
(in accuracy) SAT or Mattsson et al.'s method.
This will considerably improve the scalability properties of multi-block MPI computations. It might even be advantageous to use it
when dividing a grid block into many smaller grids to be dealt by different MPI processes as in binary black hole simulations in
 General Relativity.
In this case the traditional way of
doing it is to pass at the boundary the whole stencil needed to compute finite differences using centered operators. The accuracy
of our method implies one could just pass among the neighboring grids the values of the fields, gaining a substantial step on
scalability.

Since the information passed along the interface is a time derivative, it behaves as a scalar with respect to coordinate changes in
space \footnote{This is of course true in the case of scalar quantities. In the case we were dealing with systems of wave
equations applied to tensor quantities, some coordinate transformations are unavoidable at interfaces.}. So, its values at both
sides of the grid, namely at two different coordinate patches, should be identified without any change. By contrast, when using
 $FO$-{\sl scheme} or $SO$-{\sl Mattsson et al.'s scheme} and passing space derivatives of the fields, a coordinate transformation is needed in the generic case at
which the boundary regions represent different curvilinear coordinate patches. Thus the new scheme requires less coding and less
computation.

This new method of dealing with interfaces is not unique to second-order systems, for its underlying ideas can be applied to many cases of interest. In
particular, it can be extended to the general case of symmetric hyperbolic first-order systems. This case is under present
investigation.

\begin{acknowledgements}

We thank Luis Lehner and Manuel Tiglio for discussions, and SeCyT-UNC, CONICET, FONCyT and the Partner Group grant of the Max
Planck Institute for Gravitational Physics (Albert Einstein Institute) for financial support. O.R. thanks Perimeter Institute for
hospitality, where part of this research was carried out.

\end{acknowledgements}

\bibliographystyle{unsrtnat}
\bibliography{PCIR_JSC_I}   


\end{document}